\documentclass[useAMS,usenatbib]{mnras}

\usepackage[utf8]{inputenc}
\usepackage{graphicx}
\usepackage{color}
\usepackage{booktabs}
\usepackage[nolist]{acronym}
\usepackage{amssymb}
\usepackage[]{natbib}
\usepackage{float}
\hypersetup{draft}
\usepackage{hyperref}

\newcommand{\eagle}{{\sc eagle}}
\def\apj{ApJ}
\def\apjs{ApJS}
\def\apjl{ApJ}

\def\mnras{MNRAS}

\def\nat{Nature}
\def\araa{ARA\&A}
\def\aap{A\&A}

\def\km{{\rm\thinspace km}}

\def\pc{{\rm\thinspace pc}}
\def\kpc{{\rm\thinspace kpc}}
\def\Gpc{{\rm\thinspace Gpc}}

\def\Mpc{{\rm\thinspace Mpc}}

\def\Msun{\hbox{$\rm\thinspace M_{\odot}$}}

\def\s{{\rm\thinspace s}}       
\def\yr{{\rm\thinspace yr}}

\def\Gyr{{\rm\thinspace Gyr}}

\def\Msunyr{\Msun\yr^{-1}}
\def\Msunpc2{{\Msun\pc}^{-2}}
\def\Msunyrkpc2{{\Msun\yr^{-1}\kpc}^{-2}}

\def\magarcsec2{{\rm\thinspace mag\thinspace arcsec}^{-2}}

\begin{document}

\title[Host galaxies of merging compact objects]{Host galaxies of merging compact objects: mass, star formation rate, metallicity and colours}
\author[Artale et al.]{M. Celeste Artale$^{1}$\thanks{E-mail:Maria.Artale@uibk.ac.at, mcartale@gmail.com}, Michela Mapelli$^{1,2,3,4}$, 
Nicola Giacobbo$^{2,3,4}$, Nadeen B. Sabha$^{1}$, \newauthor Mario Spera$^{1,2,3,4,5,6}$, Filippo Santoliquido$^2$ \& Alessandro Bressan$^{7}$\\
$^{1}$Institut f\"{u}r Astro- und Teilchenphysik, Universit\"{a}t Innsbruck, Technikerstrasse 25/8, 6020 Innsbruck, Austria\\
$^{2}$Physics and Astronomy Department Galileo Galilei, University of Padova, Vicolo dell'Osservatorio 3, I--35122, Padova, Italy\\
$^{3}$INAF-Osservatorio Astronomico di Padova, Vicolo dell'Osservatorio 5, I--35122, Padova, Italy\\
$^{4}$INFN-Padova, Via Marzolo 8, I--35131 Padova, Italy\\
$^{5}$Center for Interdisciplinary Exploration and Research in Astrophysics (CIERA), Evanston, IL 60208, USA \\
$^{6}$Department of Physics \& Astronomy, Northwestern University, Evanston, IL 60208, USA \\
$^{7}$Scuola Internazionale Superiore di Studi Avanzati (SISSA), Via Bonomea 265, I-34136 Trieste, Italy\\
}
% \pagerange{\pageref{firstpage}--\pageref{lastpage}} \pubyear{2015}
\maketitle
\begin{abstract}
Characterizing the properties of the host galaxies of merging compact objects provides essential clues to interpret current and future gravitational-wave detections.
  Here, we investigate the stellar mass, star formation rate (SFR), metallicity and colours of the host galaxies of merging compact objects in the local Universe, by combining the results of {\textsc{mobse}} population-synthesis models together with galaxy catalogs from the \eagle\ simulation.  We predict that the stellar mass of the host galaxy is an excellent tracer of the merger rate per galaxy ${\rm n}_{\rm GW}$ of double neutron stars (DNSs), double black holes (DBHs) and black hole-neutron star binaries (BHNSs). We find a significant correlation also between ${\rm n}_{\rm GW}$ and SFR. As a consequence, ${\rm n}_{\rm GW}$ correlates also with the $r-$band luminosity and with the $g-r$ colour of the host galaxies. Interestingly, $\gtrsim{}60$ \%, $\gtrsim{}64$~\% and $\gtrsim{}73$~\% of all the DNSs, BHNSs and DBHs merging in the local Universe lie in early-type  galaxies, such as NGC~4993. We predict a local DNS merger rate density of $\sim{}238$ Gpc$^{-3}$ yr$^{-1}$ and a DNS merger rate $\sim{}16-121$ Myr$^{-1}$  for Milky Way-like galaxies. Thus, our results are consistent with both the DNS merger rate inferred from GW170817 and the one inferred from Galactic DNSs.
\end{abstract}


\section{Introduction}

The direct detection of gravitational waves (GWs) from  LIGO \citep{LIGOdetector} and Virgo \citep{VIRGOdetector} has opened a new perspective to investigate the nature of merging compact objects.
Eleven GW sources have been detected so far. Ten of them are identified as double black holes \citep[DBHs,][]{Abbott2016a,Abbott2016b,Abbott2016c,Abbott2016d,Abbott2017,AbbottO2a,AbbottO2b}, 
and one as a double neutron star \citep[DNS,][]{Abbott2017b,Abbott2017c}. 
The number of detections is expected to increase significantly in the near future, with the third observing run, and black hole--neutron star systems (BHNS) might be also detected \citep{Giacobbo2018B,Mapelli2018b}.

Several mechanisms have been proposed to explain the formation of merging black holes (BHs) and neutron stars (NSs). The evolution of massive stellar binaries in the field can lead to the formation of merging compact objects, especially if driven by a common envelope phase \citep{Tutukov1973,Flannery1975, Bethe1998, Portegieszwart1998,  Portegieszwart2000,Belczynski2002,Perna2002,Voss2003, Podsiadlowski2004,Podsiadlowski2005,Belczynski2006,Belczynski2007,Belczynski2008,Bogomazov2007,Moody2009,Dominik2012,Dominik2013,Dominik2015,  Mapelli2013, Mennekens2014,Tauris2015, Tauris2017,Belczynski2016,Chruslinska2018,Mapelli2017,Giacobbo2018,Giacobbo2018B,Giacobbo2018c,Mapelli2018,Mapelli2018b,Kruckow2018,Shao2018,Eldridge2016,Stevenson2017,Spera2019} or by chemically homogeneous evolution \citep{Demink2015,Demink2016,Marchant2016}. Alternative evolutionary channels are dynamical formation in star  clusters \citep{Sigurdsson1993,Kulkarni1993,Sigurdsson1995,Portegieszwart2000,Colpi2003,Oleary2006,Sadowski2008,Oleary2009,Downing2010,Downing2011,Banerjee2010,Mapelli2013,Mapelli2014,Ziosi2014,Rodriguez2015,Rodriguez2016,Antonini2016,Mapelli2016,Kimpson2016,Hurley2016,Oleary2016,Askar2017,Askar2018,Zevin2017,Banerjee2017,Antonini2018,Banerjee2018,Rastello2018,Rodriguez2018,Samsing2018,kumamoto2018,DiCarlo2019} or in galactic nuclei \citep{Oleary2009,McKernan2012,Bartos2017,Kelley2017,Stone2017,rasskazov2019}.

The sky localisation of GW sources and the characterisation of their host galaxies are crucial to identify the most likely formation mechanism.
Given current uncertainty on sky localisation inferred from GW data (still of the order of tens of square degrees in the best case, see e.g. \citealt{Abbott2018}), the detection of an electromagnetic counterpart associated with the GW event is essential to localise the host galaxy.

The spectacular detection of the counterpart of GW170817, sweeping almost the entire electromagnetic spectrum from radio to gamma-ray wavelengths \citep{abbottmultimessenger,abbottGRB,goldstein2017,savchenko2017,margutti2017,coulter2017,soares-santos2017,chornock2017,cowperthwaite2017,nicholl2017,pian2017,alexander2017} has lead to the unique identification of the host galaxy, NGC4993. This is an early-type galaxy at redshift $z\sim{}0.009783$ \citep{Levan2017}, which probably underwent a major merger recently, with stellar mass of $0.3-1.2\times10^{11}\Msun$ and metallicity in the range of 20\%-100\% the solar metallicity \citep{Im2017}.

In contrast, none of the GW events interpreted as DBH mergers was associated with an electromagnetic detection and thus their host galaxies were not identified. Hence, theoretical models are required to investigate the environment of merging DBHs and BHNSs.

A way to explore the nature of host galaxies of merging compact objects is by combining galaxy formation models with binary population synthesis. With this approach, we can reconstruct the properties of simulated host galaxies (e.g. mass, star formation rate, galaxy type) and possibly we can inform the low-latency search for electromagnetic counterparts \citep{delpozzo2018} or, even if the electromagnetic counterpart is not observed, we can infer astrophysically motivated criteria to localise the most likely host galaxy of a GW event \citep{Mapelli2018}.

The main challenge of this approach is the extreme physical range between galaxy formation and compact object binary formation. Several works have attempted to  address this issue by (semi)analytical models or by cosmological simulations \citep[e.g.,][]{OShaughnessy2010,Belczynski2016,Lamberts2016,Dvorkin2016,OShaughnessy2017,Schneider2017,Cao2018,Elbert2018,Lamberts2018,Mapelli2018,Perna2018,Eldridge2019,Marassi2019,mapelli2019,Safarzadeh2019}.  

Using a sample of zoom-in simulations, \citet{OShaughnessy2017} discuss the impact of the assembly history of galaxies on DBH mergers and suggest that DBHs are more likely to form in nearby metal-poor dwarf galaxies. 
\citet{Cao2018} make an exhaustive analysis of the host galaxies of DBHs by using the cosmological N-body simulation Millennium-II with semi-analytical galaxy formation recipes from \citet{Guo2011}. Assuming different delay times for DBHs, they find that at $z=0.3$
massive merging DBHs (with total mass $\gtrsim 50\Msun$) formed at high redshift reside in galaxies with stellar mass $\sim 4.7\times10^{10}\Msun$, while DBHs formed recently are located in $\sim 10^{7}-10^{9}\Msun$ galaxies. 

\citet{Lamberts2018} investigate DBH systems in  Milky Way-type galaxies, combining zoom--in simulations with population synthesis models. Interestingly, they find that one third of the DBHs were formed ex-situ of the main galaxy, in satellite galaxies that eventually merged.

Combining the results of population synthesis simulations
with the cosmological box Illustris--1 \citep{Vogelsberger2014}, \citet{Mapelli2018} explore the host galaxies of merging DNSs, DBHs and BHNSs in the local Universe. They find that DNSs tend to form and merge in galaxies with stellar mass $10^{9}-10^{12}\Msun$ with short delay times.
In contrast, BHNSs and DBHs form preferentially in low mass galaxies ($<10^{10}\Msun$), but merge either in massive or low mass galaxies with longer delay times than DNS sources. These results originate from the different metallicity dependence of merging compact objects: DBHs and BHNSs are expected to form more efficiently from metal-poor progenitors, while DNSs form almost independently of progenitor's metallicity \citep{Giacobbo2018B}.

In this work, we investigate the properties of the host galaxies of merging DBHs, DNSs and BHNS in the local Universe. 
We combine the galaxy catalogs from the hydrodynamical cosmological simulation \eagle\ \citep{Schaye15} with the catalogs of merging compact objects from the population synthesis code {\sc mobse} \citep{Giacobbo2018}.

Our methodology is similar to the one presented by \citet{Mapelli2017,Mapelli2018}, but we adopt a cosmological box with a resolution $\sim 5.5$ higher.
This enables us to probe host galaxies to lower stellar mass (down to $\sim{}10^7$ M$_\odot$), which are unresolved in the Illustris-1 simulation adopted by \citet{Mapelli2017}. Moreover, the sub-grid physical models adopted in the \eagle{} and in the Illustris simulations are drastically different. Thus, we also want to understand the uncertainties introduced in our work by the choice of a cosmological simulation.

This paper is organised as follows. In \S~\ref{sec:simulations} we describe the main properties of \eagle\ and {\sc mobse}, while the methodology is explained in \S~\ref{sec:method}.
We investigate the connection between the GW sources and the star formation rate, stellar mass and metallicity of the host galaxies in \S~\ref{sec:results}. In \S~\ref{sec:discussion}, we discuss the implications of our results on the merger rate in late-type and early-type galaxies. Our main conclusions are discussed in \S~\ref{sec:conclusions}.

\section{Methods}\label{sec:simulations}

We compute the number of DNSs, DBHs and BHNSs by combining the results from the population synthesis code 
{\sc mobse}  \citep{Giacobbo2018} with galaxy catalogs from the hydrodynamical cosmological
simulation {\sc eagle}. In this section, we present the general features of these codes. 

\subsection{Population synthesis code: {\sc mobse}}

The population synthesis code {\sc mobse}  \citep{Giacobbo2018} represents an upgrade of the 
{\sc bse} code \citep{Hurley2000,Hurley2002} including new prescriptions for stellar winds \citep[see][]{Vink2001,Vink2005,Chen2015}, core collapse supernovae (SNe) based on \citet{Fryer2012}, and pair-instability and pulsational pair-instability SNe \citep{Woosley2017,Spera2017}.

In particular, {\sc mobse} describes the mass loss by stellar winds of massive hot stars (O- and B-type main sequence stars, Wolf-Rayet stars and luminous blue variable stars)  as $\dot{M}\propto{}Z^{\beta{}}$, where $\beta{}$ is defined in the following way:
\begin{equation}
\label{eq:scaling}
\beta = \left\{ \begin{array}{lr}
  0.85 & \rm{if}~~~ \Gamma_{\rm e} < 2/3 \\
  2.45 - 2.40 \,{} \Gamma_{\rm e} & \rm{if}~~~ 2/3 \leq \Gamma_e \leq 1~ \\
  0.05\,{} & \rm{if}~~~ \Gamma_{\rm e}>1.\end{array}\right.
\end{equation}
Here $\Gamma_{\rm e}$ is the electron-scattering Eddington ratio, expressed as (see eq.~8 of \citealt{Graefener2011}):
\begin{equation}\label{eq:gamma}
\log{\Gamma_e}=-4.813+\log{(1+X_{\rm H})}+\log{(L/L_\odot)}-\log{(M/\Msun)}.
\end{equation}
In equation~\ref{eq:gamma}, $X_{\rm H}$ is the Hydrogen fraction, $L$ is the star luminosity and $M$ is the star mass.

Accounting for the dependence of mass loss on both metallicity ($Z$, \citealt{Vink2001}) and Eddington ratio \citep{Vink2016} is a key ingredient to predict the final mass of a compact object \citep{mapelli2009,mapelli2010,belczynski2010,Mapelli2013,spera2015}. In fact, massive metal-poor stars are predicted to lose much less mass by stellar winds than metal-rich stars, ending their life with larger cores and larger envelopes. This implies that massive metal-poor stars are more likely to directly collapse into BHs, without an SN explosion, giving birth to more massive remnants than metal-rich stars.

In {\sc mobse}, the final mass of the compact object depends on the Carbon-Oxygen core mass and on the final mass of the progenitor star as described in \citet{Fryer2012}. In particular, in this paper we make use of the rapid core-collapse SN model described by
\citet{Fryer2012}. {\sc mobse} also includes a description of electron-capture SNe, as detailed in \cite{Giacobbo2018c}. Massive stars developing a Helium core between 32 and 64 M$_\odot$ are expected to develop pulsational pair instability, which is modelled as described in \citet{Spera2017}. Finally, stars with Helium core in the $64-135$ M$_\odot$ range are expected to undergo pair instability and to be completely disrupted, leaving no compact object.

\begin{figure}
\centering
\includegraphics[width=0.46\textwidth]{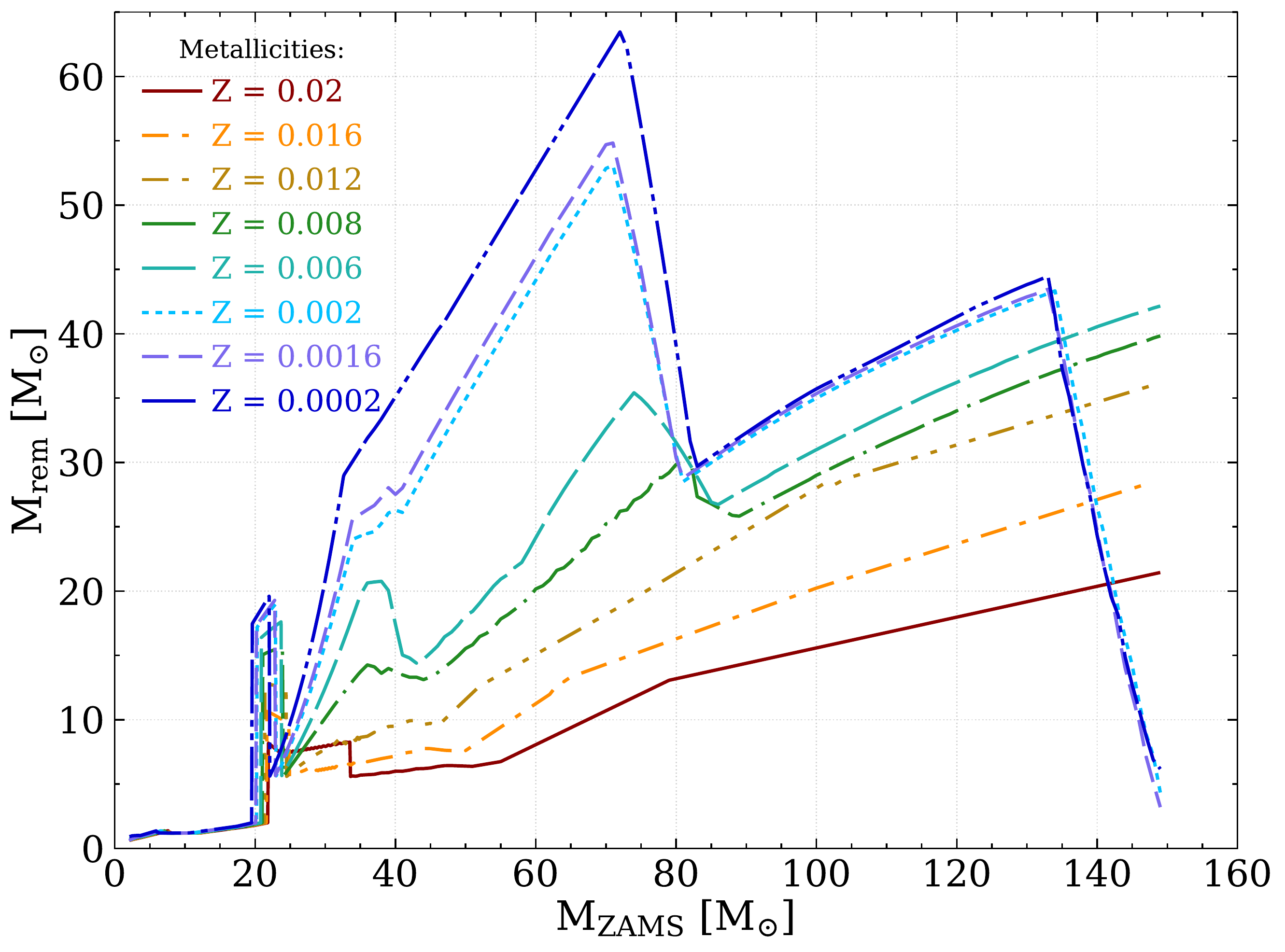}
\caption{Mass of the compact object (M$_{\rm rem}$) as a function of the mass of the progenitor star (M$_{\rm ZAMS}$) for 8
metallicities between $Z=0.02$ and $0.0002$. See \citet{Giacobbo2018} for details on this model.}
\label{fig:MOBSE}
\end{figure}

These prescriptions produce a mass spectrum of compact objects as shown in Figure~\ref{fig:MOBSE}. The mass distribution shows a dearth of compact objects with mass between $\sim{}2$ and $\sim{}5$ M$_\odot$, consistent with the mass gap between NSs and BHs suggested by \citet{ozel2010} and \citet{farr2011}. The maximum mass of BHs in this model is $\sim{}65$ M$_\odot$ (at metallicity $Z=0.0002$), although \citet{Giacobbo2018} and \citet{Giacobbo2018B} show that only BHs with mass $\lesssim{}40$ M$_\odot$ are able to merge by GW emission within a Hubble time.

%%%%%%%%%%%%%%%%%%%%%%%%%%%%%%%%%%%%FIGURE 1%%%%%%%%%%%%%%%%%%%%%%%%%%%%%%%%
%%%%%%%%%%%%%%%%%%%%%%%%%%%%%%%%%%%%%%%%%%%%%%%%%%%%%%%%%%%%%%%%%%%%%%%%%%%%

In this work, we use the catalog of merging compact objects corresponding to the run named as  CC15$\alpha{}$5
from \citet{Giacobbo2018B}, also implemented in \citet{Mapelli2018}.
This run matches the cosmic merger rate reported by recent LIGO-Virgo results \citep[see Figure 15 in Giacobbo \&{} Mapelli 2018]{Giacobbo2018B}. It also matches the expected merger rate of DNSs in the Milky Way, according to the estimate by \cite{Pol2019}. In this run we adopt high efficiency of common-envelope ejection ($\alpha = 5$) and low SN kicks using a Maxwellian curve with root mean square $\sigma = 15 \km \s^{-1}$. Low natal kicks are consistent also with the orbital properties of some DNSs in the Milky Way \citep{heuvel2007,beniamini2016}, with the presence of r-process material in ultra-faint dwarf galaxies \citep{beniamini2016b} and with several families of X-ray binaries (e.g. \citealt{pfahl2002,knigge2011,Tauris2017}, and references therein).

The simulation is composed of 12 sub-sets at different metallicities $Z = 0.0002$, 0.0004, 0.0008,
0.0012, 0.0016, 0.002, 0.004, 0.006, 0.008, 0.012, 0.016 and 0.02.
We adopt as solar metallicity $Z_{\odot} = 0.02$.
For each sub-set we simulate $10^7$ stellar binaries, hence the total number of binaries is $1.2\times10^{8}$.

We combine the catalog of merging compact objects obtained with {\sc mobse} with the galaxy catalog from the cosmological hydrodynamical simulation \eagle. 
We describe the main properties of \eagle{} in the following section.

\subsection{The EAGLE simulation}

The \eagle\ simulation suite \citep[][]{Schaye15,Crain15} is a set of cosmological hydrodynamical simulations 
that were run with a modified version of {\sc gadget-3} code.
In this work we use the simulation labeled as RecalL0025N0752 available on the {\sc SQL} 
database \citep[see][]{McAlpine2016}\footnote{{\url http://icc.dur.ac.uk/Eagle/}, {\url http://virgo.dur.ac.uk/data.php}}. 
This run represents the highest resolution available of \eagle\ suite, and at the same time, contains a statistical sample of galaxies with different morphologies, stellar masses and star formation rates. 

The RecalL0025N0752 run represents a periodic box of $25\Mpc$ side that initially contains $752^3$ gas and dark matter particles
with masses of $m_{\rm gas} = 2.26\times10^5 \Msun$ and $m_{\rm DM} = 1.21\times10^6 \Msun$. Henceforth, we refer to this run simply as \eagle.
The simulation was run from $z=127$ up to $z\sim 0$, adopting 
the $\Lambda$CDM cosmology with parameters inferred from \cite{Planck13} ($\Omega_{\rm m} = 0.2588$, $\Omega_\Lambda = 0.693$,
$\Omega_{\rm b} = 0.0482$, and $H_0 = 100\; h$ km s$^{-1}$ Mpc$^{-1}$ with $h = 0.6777$). 
We have selected the RecalL0025N0752 EAGLE simulation since it enables us to investigate host galaxies with a lower stellar mass ($10^7-10^8 \Msun$) not well resolved in the larger (100 Mpc) \eagle\ box.

The \eagle\ simulation includes different subgrid models to account for different processes behind galaxy formation such as
star formation, radiative cooling and heating, stellar evolution, chemical enrichment, UV/X-ray ionizing background, 
AGB stars and SN feedback, and AGN feedback.
For this run, the parameters of the subgrid models are calibrated to reproduce the galaxy sizes at $z\sim0$, the observed stellar mass function, and the relation between the black hole and stellar masses \citep[see][for further details]{Schaye15}.

From the database, we can obtain the information of the stellar particles and the galaxy properties such as the total stellar mass, the metallicity of the star-forming gas, and the star formation rate.
In this work we select all galaxies above $M_\ast{} > 10^{7}\Msun$ which corresponds to have at least $\sim{}44$ stellar particles within the galaxy.

We also explore the colours of the host galaxies of DNSs by computing their colour--magnitude diagram. 
For this we use the dust attenuated absolute $g$ and $r$ bands in AB magnitudes in the rest frame of galaxies.
These magnitudes were computed using the radiative transfer code {\sc skirt} for galaxies with stellar mass $M_\ast{} > 10^{8.5}\Msun$,
which contains more than 250 dust particles. 
The details of the procedure are described in \citet{Camps2016,Camps2018}, and the magnitudes are provided in the \eagle\ database.

\subsection{Monte Carlo Method}\label{sec:method}
 
To compute the number of merging compact objects within the host galaxies in the nearby Universe, we first combine the results from the population synthesis run
CC15$\alpha{}$5 with the information of the stellar particles from \eagle. 
For this, we follow the methodology presented in \citet{Mapelli2017} (see \S~2.3 of the aforementioned work), also implemented in \citet{Mapelli2018b} and \citet{Mapelli2018}. In this section we summarise the main concepts of this method.

From the \eagle\ simulation, we first select the stellar particles formed in each snapshot as the progenitors of merging compact objects. 
We use the initial mass $m_\ast$, formation time $t_\ast$ and metallicity $Z_\ast$ of each stellar particle, and we find the metallicity that best matches $Z_\ast$ among the 12 simulated metallicities of run CC15$\alpha{}$5. 

For each sub-set of CC15$\alpha{}$5, we use the total initial stellar mass $M_{\rm MOBSE}$,
and the total number of merging compact objects $N_{\rm MOBSE}$ (corresponding to DBHs, DNSs or BHNS).
Then for each stellar particle, we combine this information and compute the number of merging compact objects ($n_{\rm mco}$)  as

\begin{equation}
 n_{\rm mco} = N_{\rm MOBSE} \frac{m_*}{M_{\rm MOBSE}} f_{\rm corr} f_{\rm bin},
\end{equation}

\noindent where the parameter $f_{\rm corr}=0.285$ is a correction factor 
accounting that the primary stars are $m_{\rm p}\geq5\Msun$, while $f_{\rm bin}=0.5$ 
is the binary fraction (we assume 50 \% of stars are in binaries).

Then, we randomly select $n_{\rm mco}$ merging compact objects from the simulated stellar population sub-set with mass $M_{\rm MOBSE}$ and we associate them with the corresponding \eagle{}  stellar particle. 
Due to the resolution of the \eagle\ suite selected in this work, most stellar particles have $n_{\rm mco}<1$ (i.e. they produce less than one merging compact object). Thus, we impose that if $n_{\rm mco}<1$ for a given particle, the next stellar particle we consider  inherits the value of $n_{\rm mco}$ from the previous particle in addition to its own value of $n_{\rm mco}$, till $n_{\rm mco}$ becomes $\ge{}1$. We assign the merging compact object to the first stellar particle for which $n_{\rm mco}\ge{}1$ and then we reset the value of $n_{\rm mco}$. 
For each merging compact object selected we save its properties such as the masses of the two compact objects and the delay time.

For each merging compact object selected as above, we estimate the look-back time of the merger, by combining the formation time of the stellar particle ($t_{\rm form}$) with the time between the formation of the progenitor binary system and the merger of the two compact-objects (delay time, $t_{\rm delay}$), as $t_{\rm merg} = t_{\rm form} - t_{\rm delay}$.
This procedure allows us to follow the position of the stellar particles
from \eagle\ simulation across cosmic time. Hence, it is possible to identify the properties of host galaxies where 
merging compact objects form and merge.

Since we want to investigate merging compact objects in the local Universe, we only select the merging compact objects corresponding to $z\lesssim0.1$ (i.e., the snapshots corresponding to $z=0$ and $z=0.1$).
Hence, in this work \textit{we investigate the properties of the host galaxies at the time compact binary systems merge}. 

We note that double compact objects can be formed at early stages of the Universe with long delay times, and merge at $z\lesssim0.1$
\citep{Cao2018,Mapelli2018}. In the following analysis, we will consider only those galaxies that contain at least one binary compact object merging at $z\lesssim0.1$ (i.e. galaxies that do not host merging compact objects are not shown).

\section{Results}\label{sec:results}

\subsection{Mass -- metallicity relation}

%%%%%%%%%%%%%%%%%%%%%%%%%%%%%FIGURE 2%%%%%%%%%%%%%%%%%%%%%%%%%%%%%%%%%%%%%%%%%
\begin{figure*}
\centering
\includegraphics[width=1.0\textwidth]{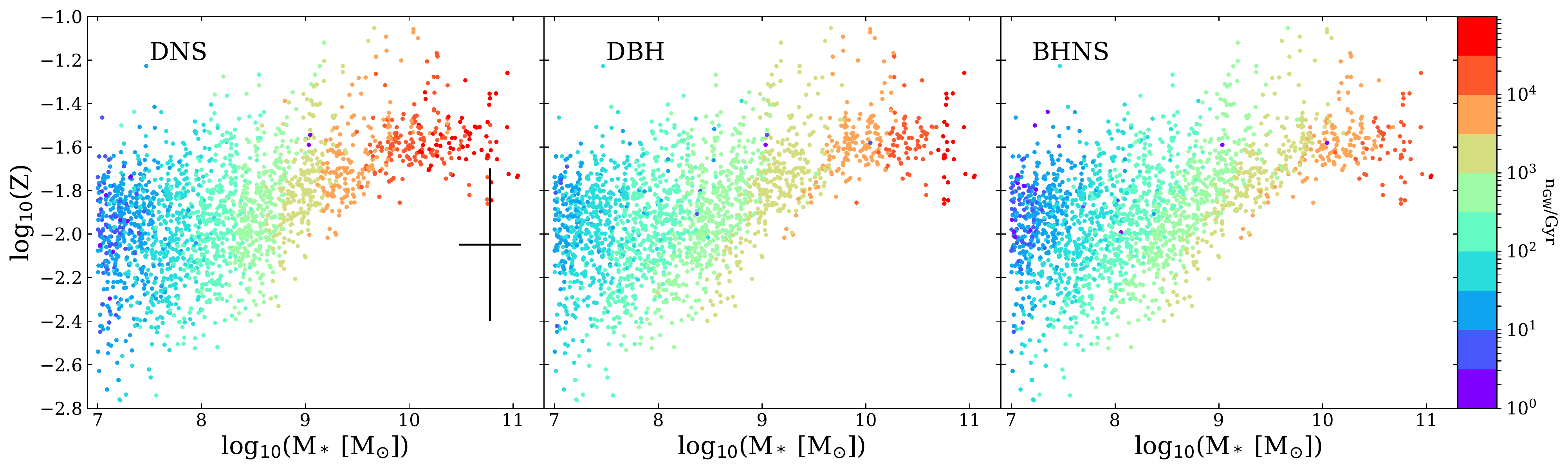}
\caption{Mass-metallicity relation of the host galaxies of merging DNSs (\textit{left panel}), DBHs (\textit{middle panel}),
and BHNSs (\textit{right panel}). Each point represents an individual galaxy from the \eagle\ catalog. The colour code represents the local merger rate per galaxy ${\rm n}_{\rm GW}$ (i.e. the total number of merging compact objects per each galaxy per redshift bin, from $z=0$ to $z=0.1$).
On the left-hand panel, the black lines represent the range of masses and metallicities computed for the galaxy NGC 4993, the host galaxy of GW170817 \citep[see,][]{Im2017}.}
\label{fig:MZR-GW25Mpc}
\end{figure*}
%%%%%%%%%%%%%%%%%%%%%%%%%%%%%%%%%%%%%%%%%%%%%%%%%%%%%%%%%%%%%%%%%%%%%%%%%%%%

Figure~\ref{fig:MZR-GW25Mpc} shows the mass-metallicity relation for the host galaxies of DNSs, DBHs and BHNSs. As expected, massive galaxies in the \eagle\ simulation are metal-rich compared with low mass galaxies, reproducing the observed trend \citep{Schaye15,DeRossi2017}.
The black lines mark the mass-metallicity range of NGC~4993, the host galaxy of GW170817 reported by \citet{Im2017}. Although the uncertainty on the metallicity of NGC~4993 is extremely large, this galaxy seems to be remarkably metal poor with respect to the average value of the galaxies in our sample. The mass of NGC~4993 matches the mass of the most massive galaxies in our sample.

The merger rate per galaxy (${\rm n}_{\rm GW}$, i.e. the number of compact object mergers per galaxy per unit time) of DNSs, BHNSs and DBHs strongly correlates with the stellar mass of the host galaxy at the time of merger. 
Interestingly, even considering the large scatter in metallicity at a given stellar mass, the stellar mass content of galaxies shows to be a fundamental tracer of the merger rate.

\subsection{Host galaxy stellar mass}~\label{sec:Results_Ms}

%%%%%%%%%%%%%%%%%%%%%%%%%%%%%%%%%%%%%FIGURE 3%%%%%%%%%%%%%%%%%%%%%%%%%%%%%%%%
\begin{figure*}
\centering
\includegraphics[width=0.45\textwidth]{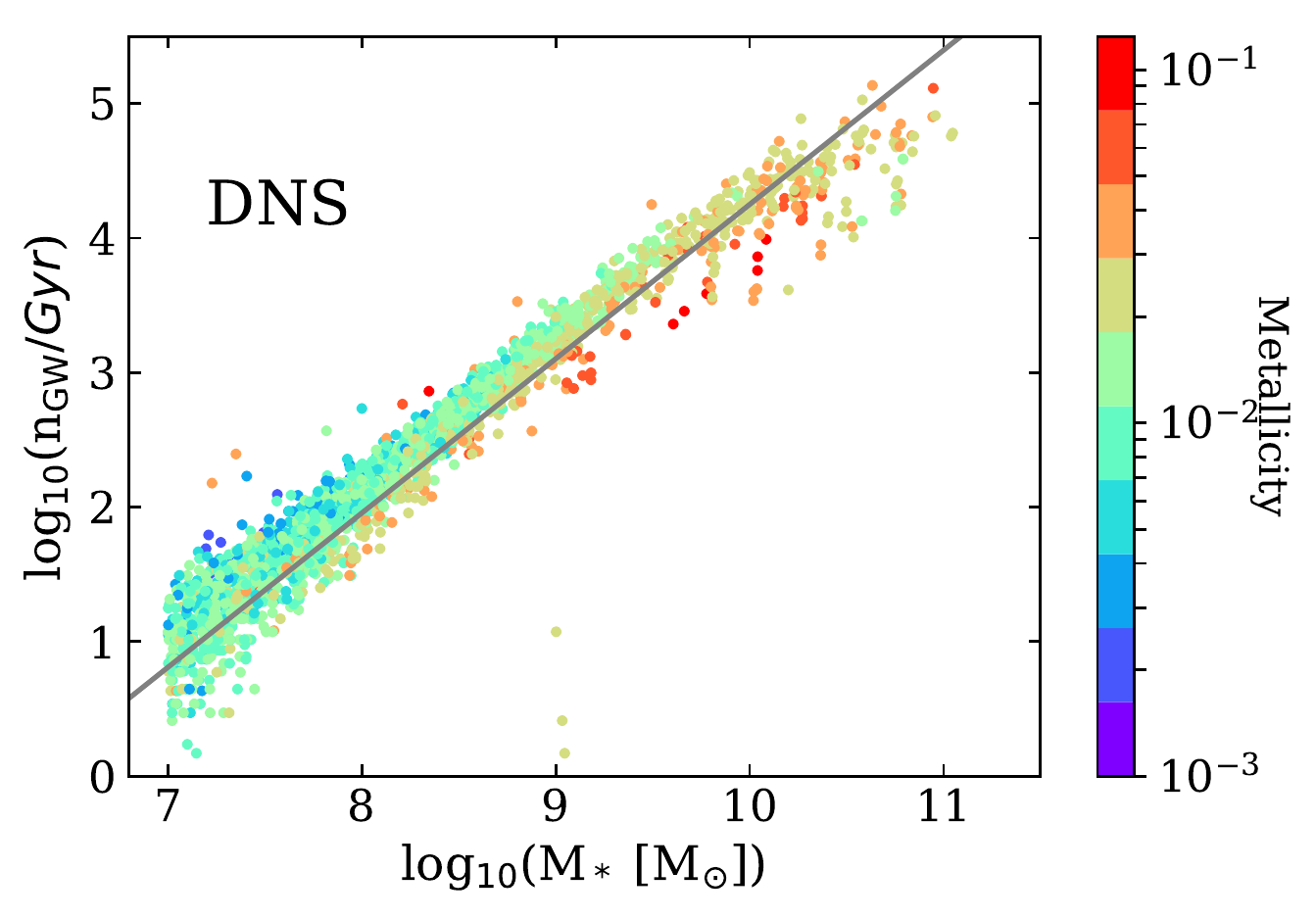}
\includegraphics[width=0.47\textwidth]{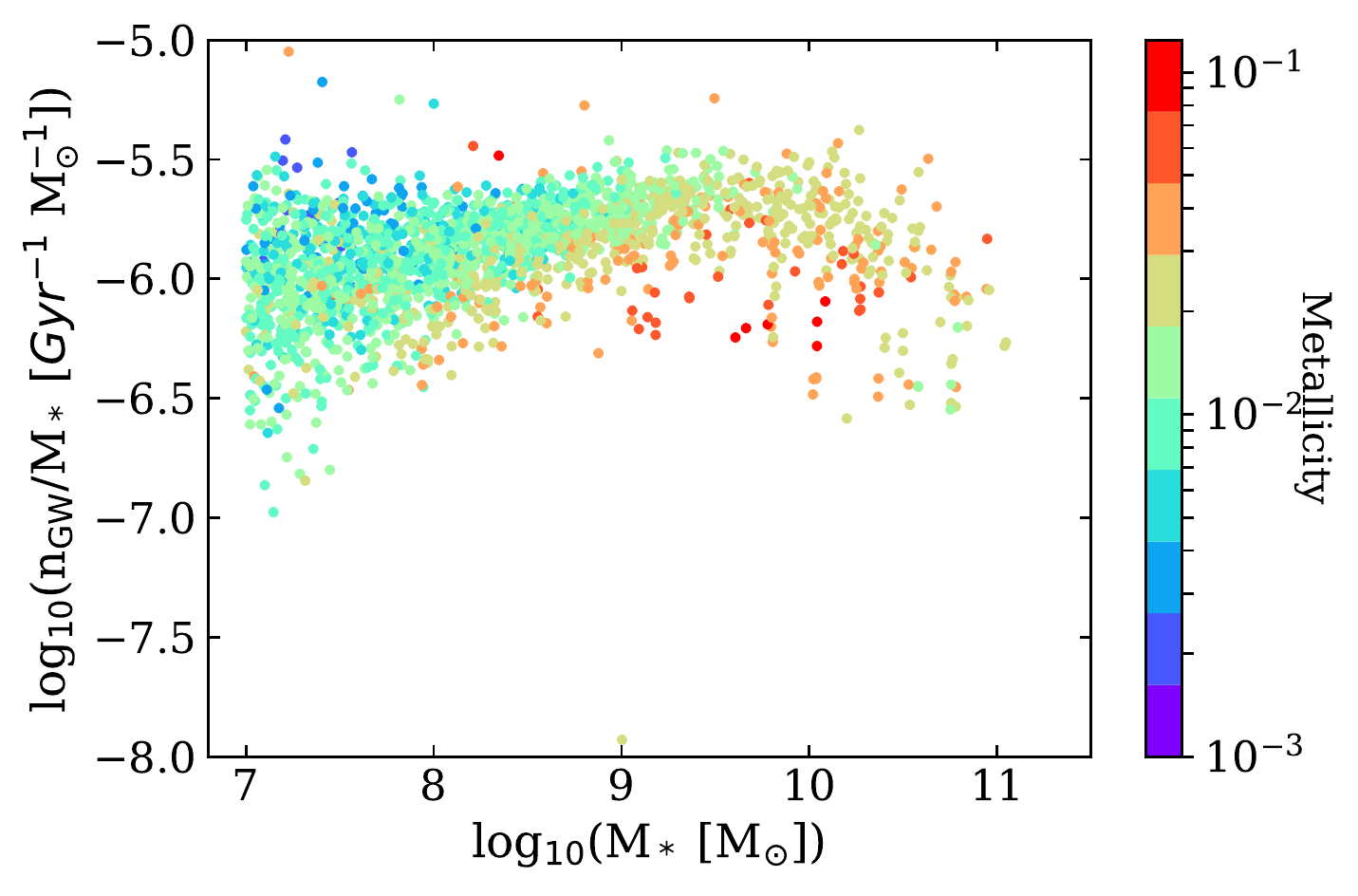}
\includegraphics[width=0.45\textwidth]{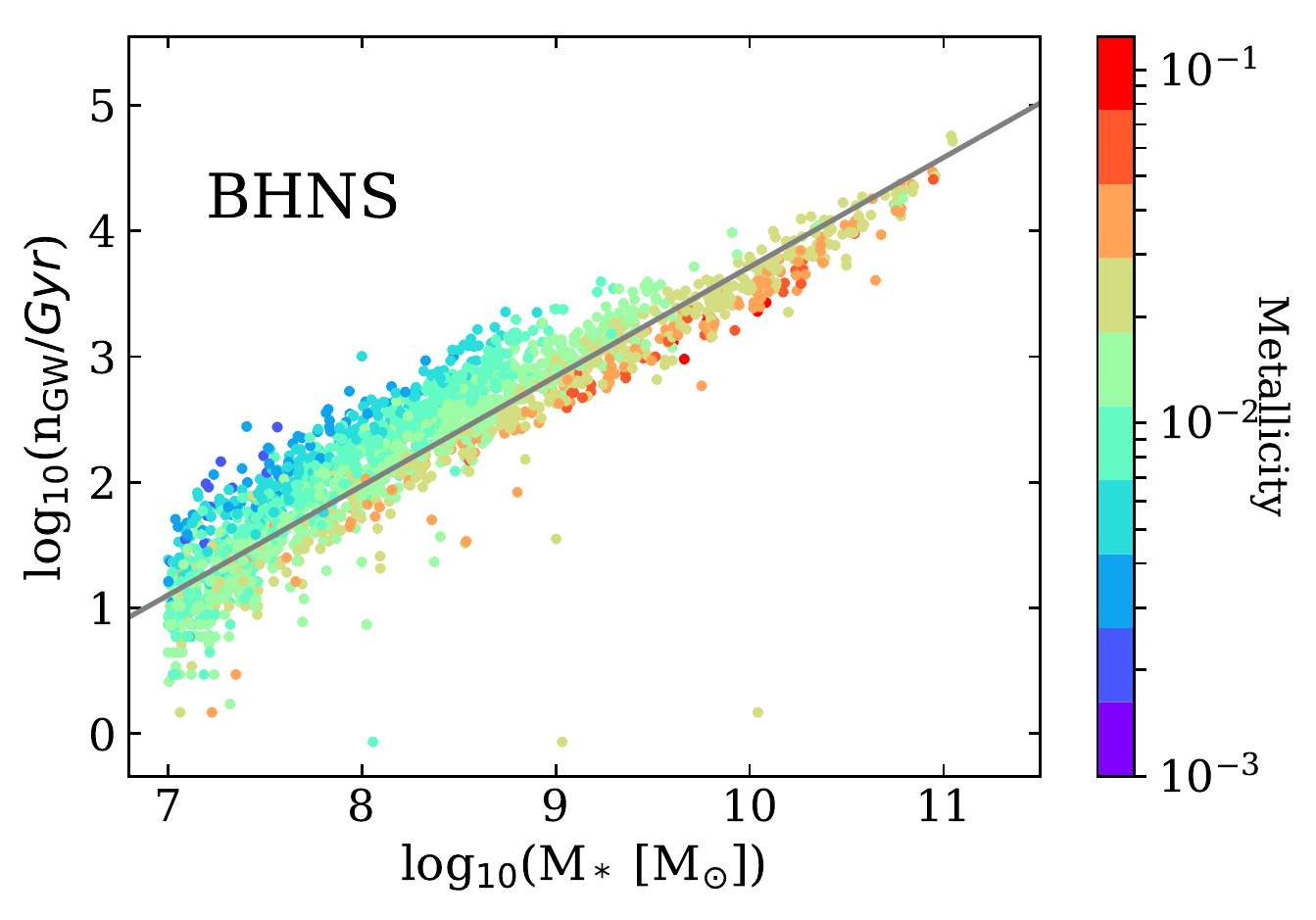}
\includegraphics[width=0.47\textwidth]{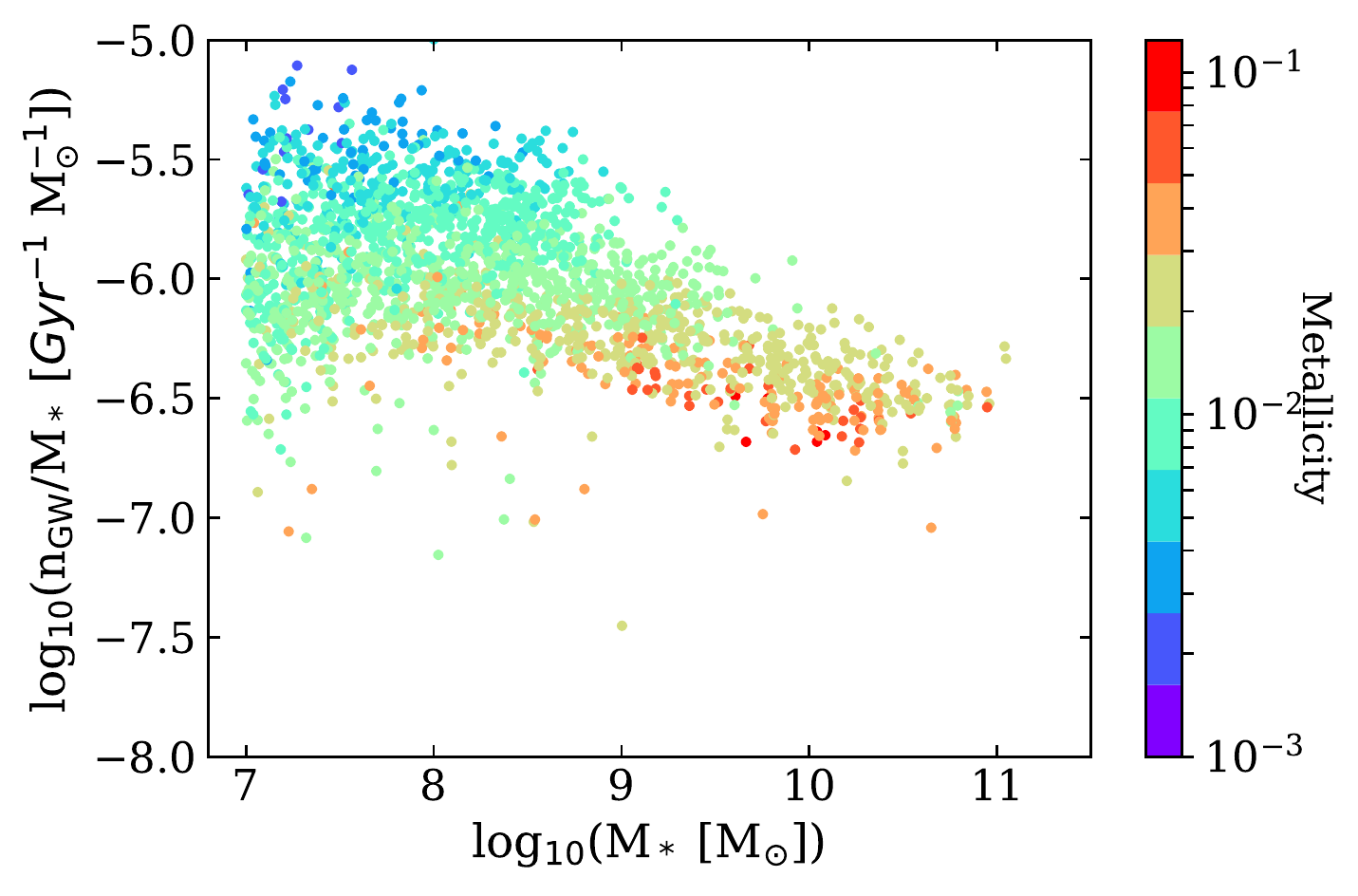}
\includegraphics[width=0.45\textwidth]{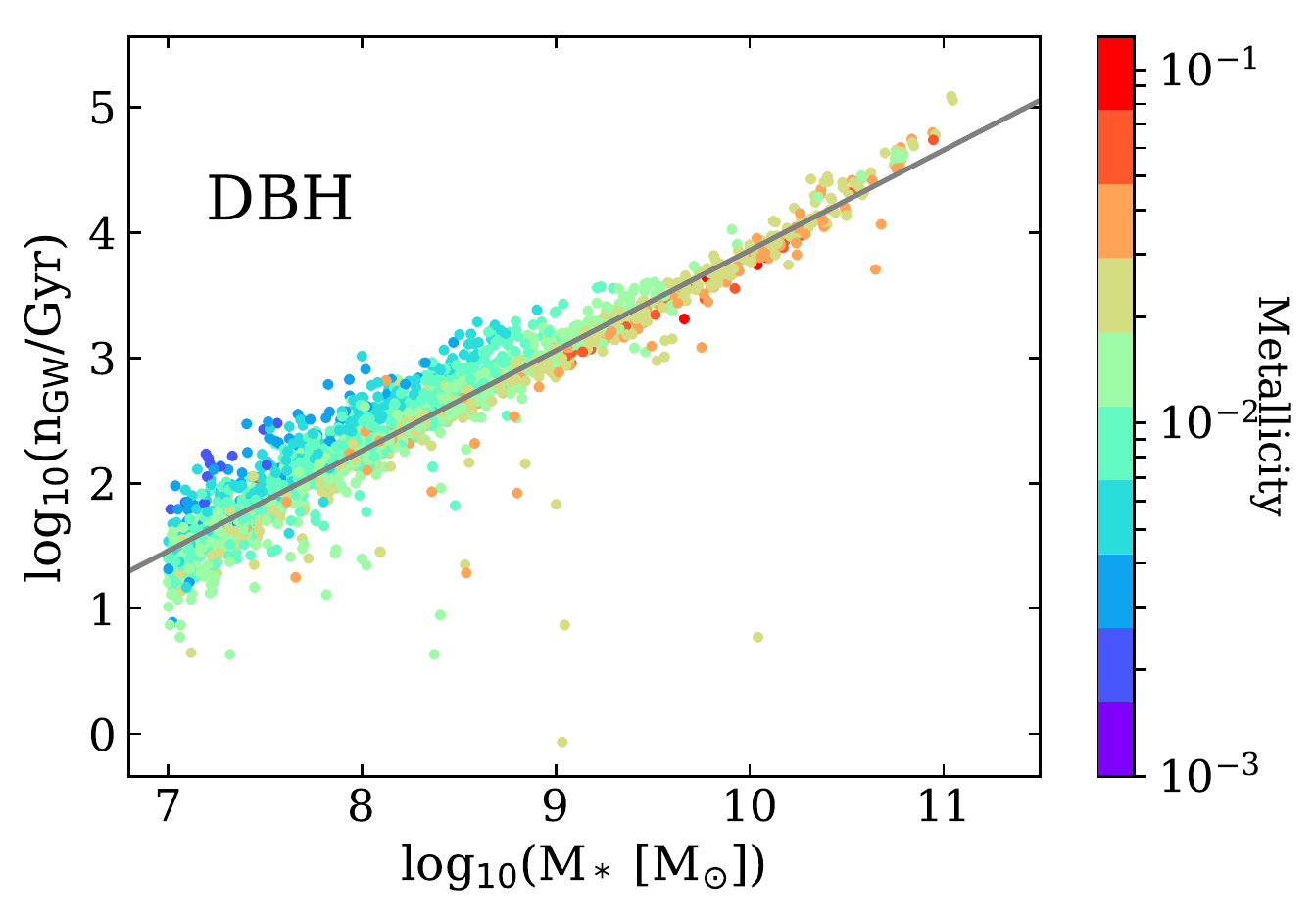}
\includegraphics[width=0.47\textwidth]{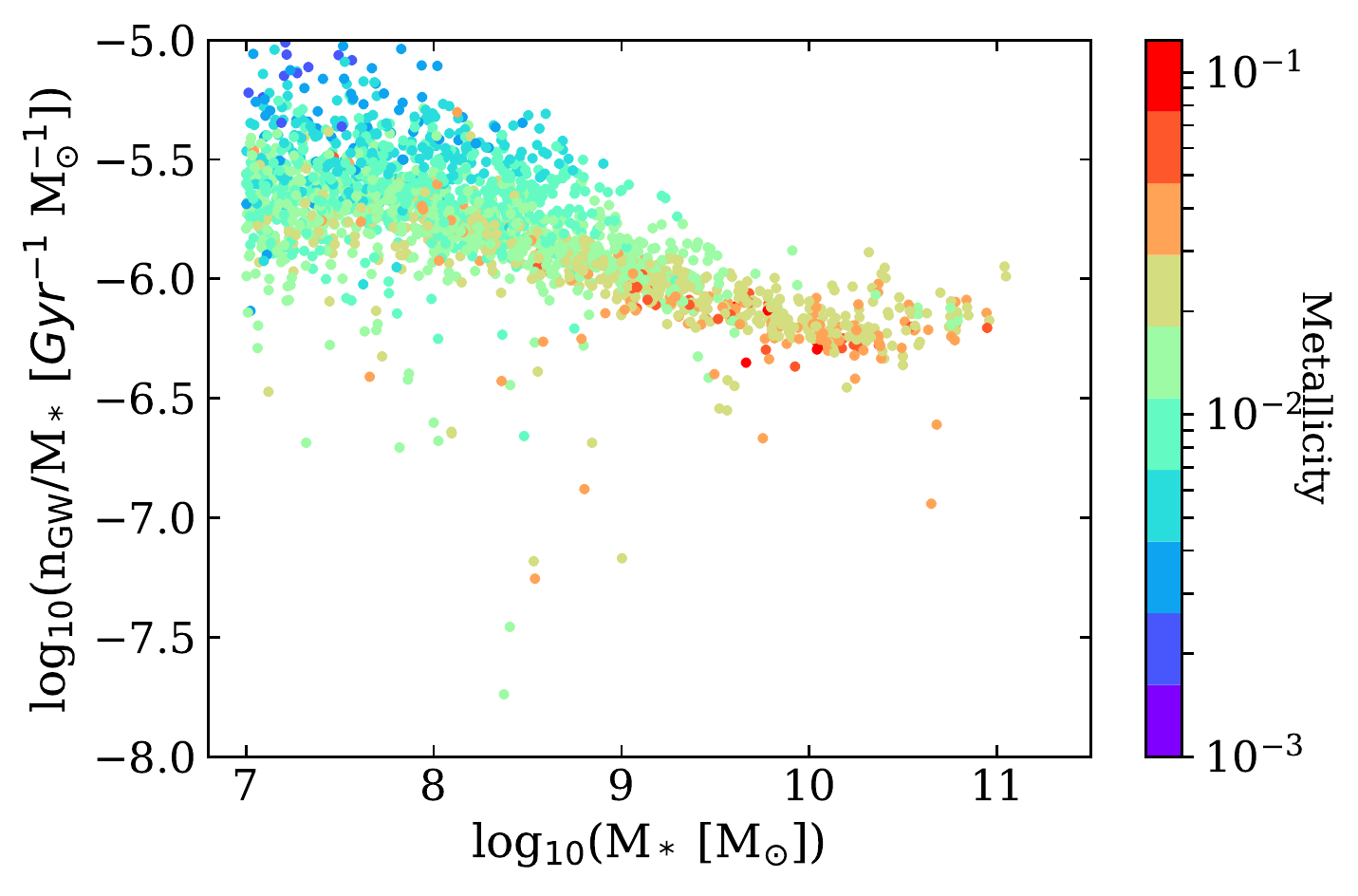}
\caption{\textit{Left-hand panel:} Merger rate per galaxy 
  as a function of the stellar mass of the host galaxy. Top panel: DNSs; middle panel: BHNSs; bottom panel: DBHs. 
We include the fits of $\log ({\rm n_{GW}}/{\rm Gyr}) = a_{\rm M_\ast{}} \log(M_\ast{}) + b_{\rm M_\ast{}}$.
\textit{Right-hand panel:} Merger rate per galaxy normalized to the galaxy mass  as a function of the stellar mass of the host galaxy.
In all the plots each point represent an individual galaxy from the {\sc eagle} catalog.
The colour code indicates the metallicity of the host galaxy.}
\label{fig:GW25Mpc-GWMs}
\end{figure*}
%%%%%%%%%%%%%%%%%%%%%%%%%%%%%%%%%%%%%%%%%%%%%%%%%%%%%%%%%%%%%%%%%%%%%%%%%%%%%%%%

%%%%%%%%%%%%%%%%%%%%%%%%%%%%%%%%%%%%%FIGURE 4%%%%%%%%%%%%%%%%%%%%%%%%%%%%%%%%
\begin{figure*}
\centering
\includegraphics[width=0.45\textwidth]{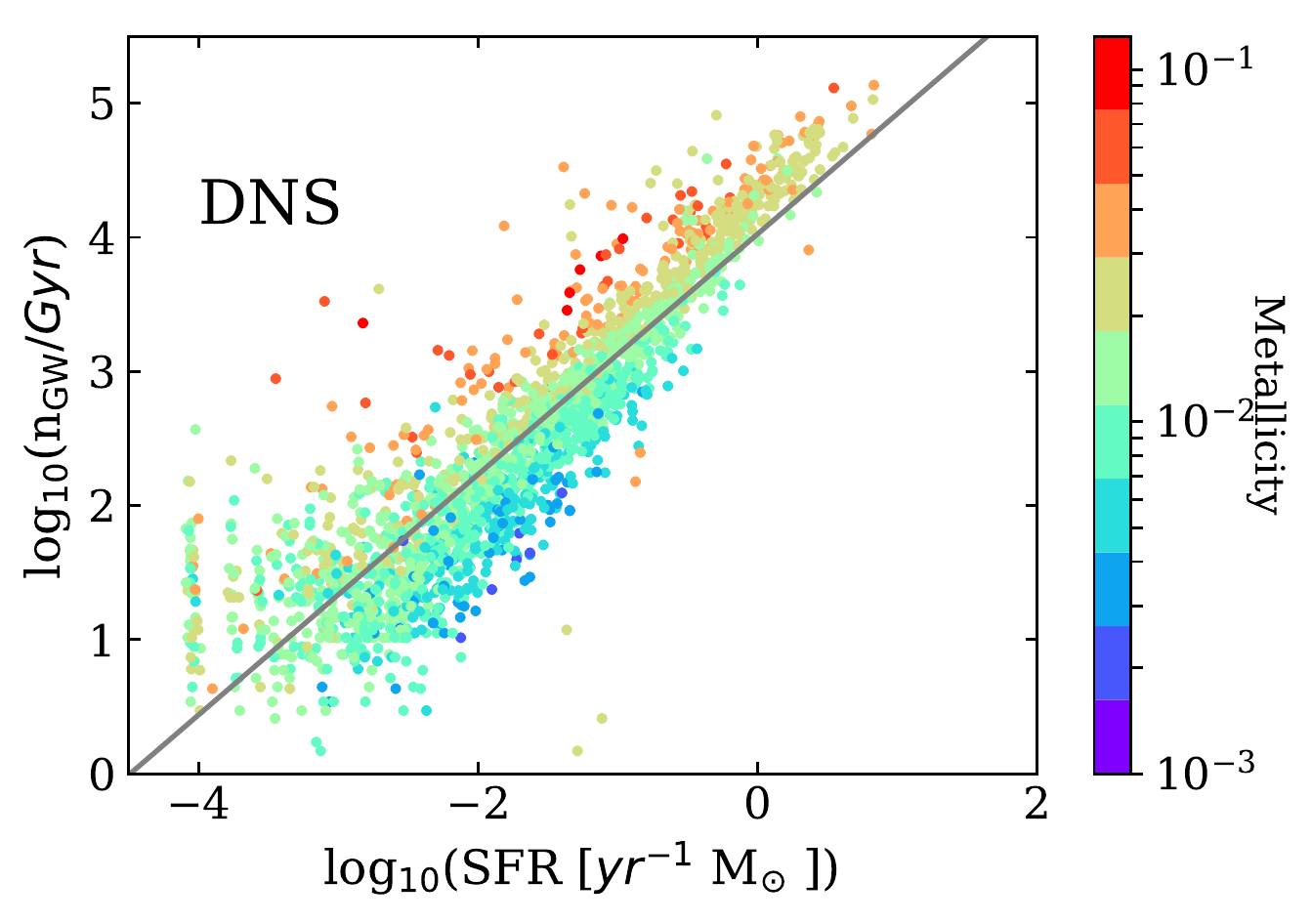}
\includegraphics[width=0.45\textwidth]{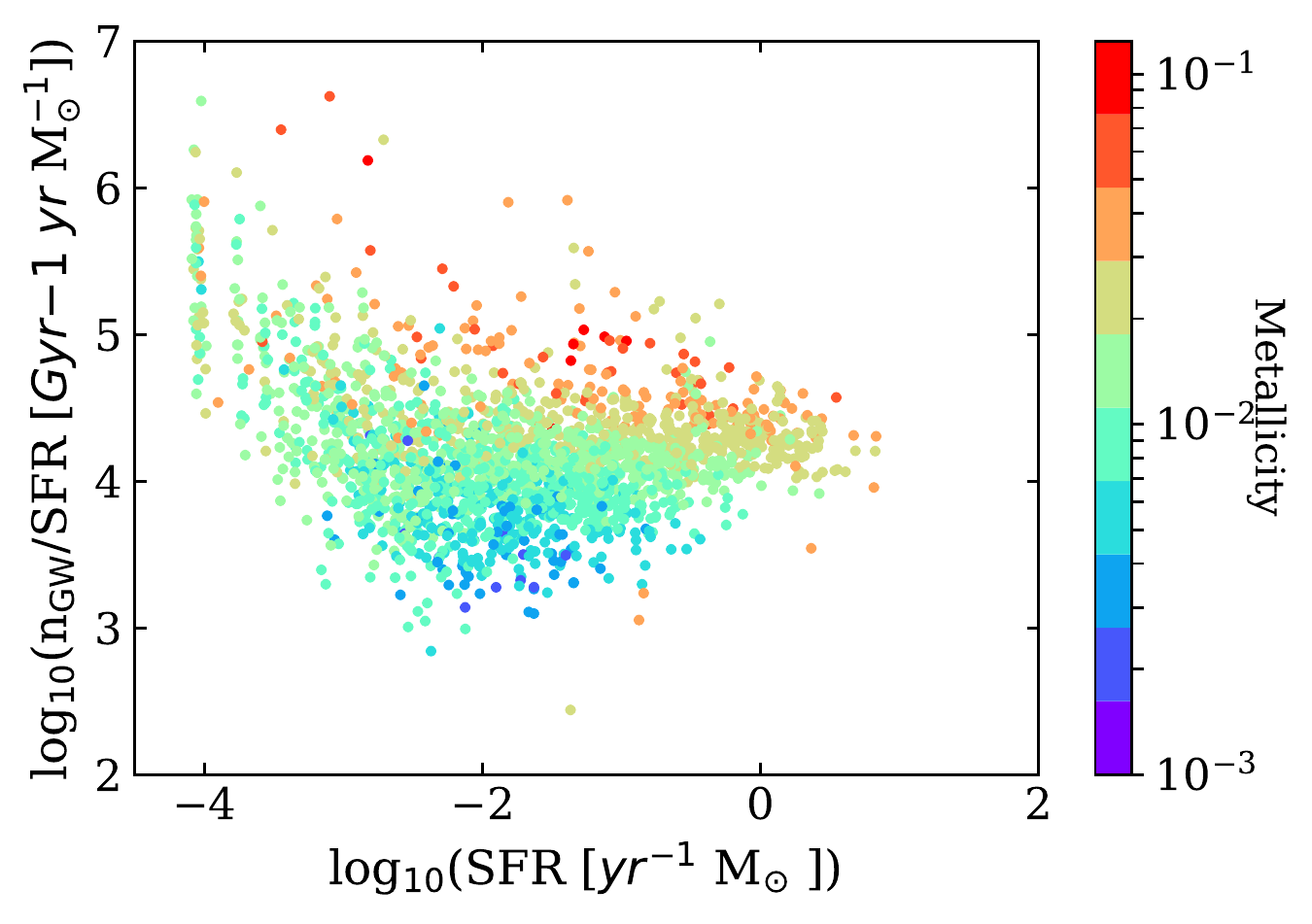}
\includegraphics[width=0.45\textwidth]{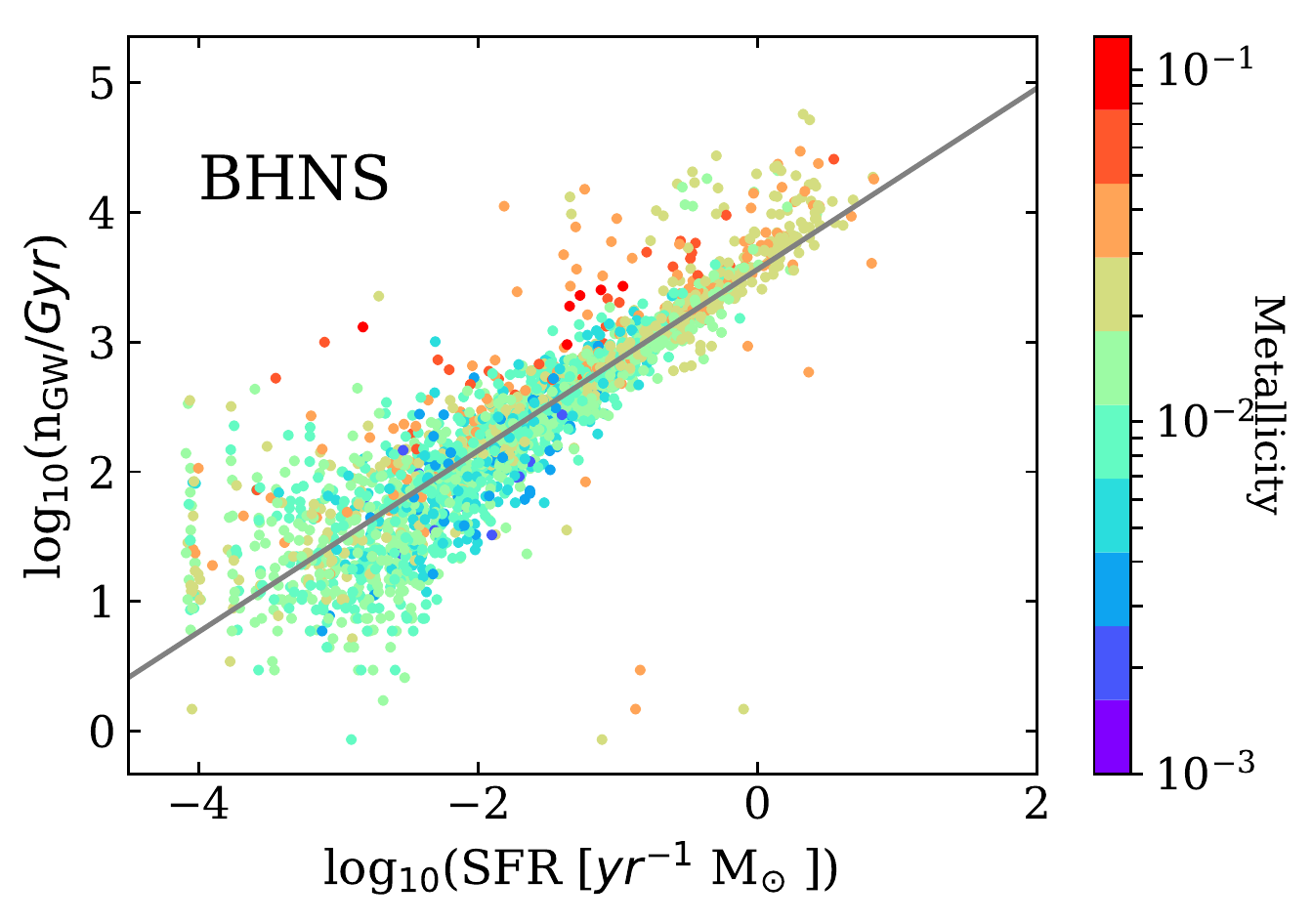}
\includegraphics[width=0.45\textwidth]{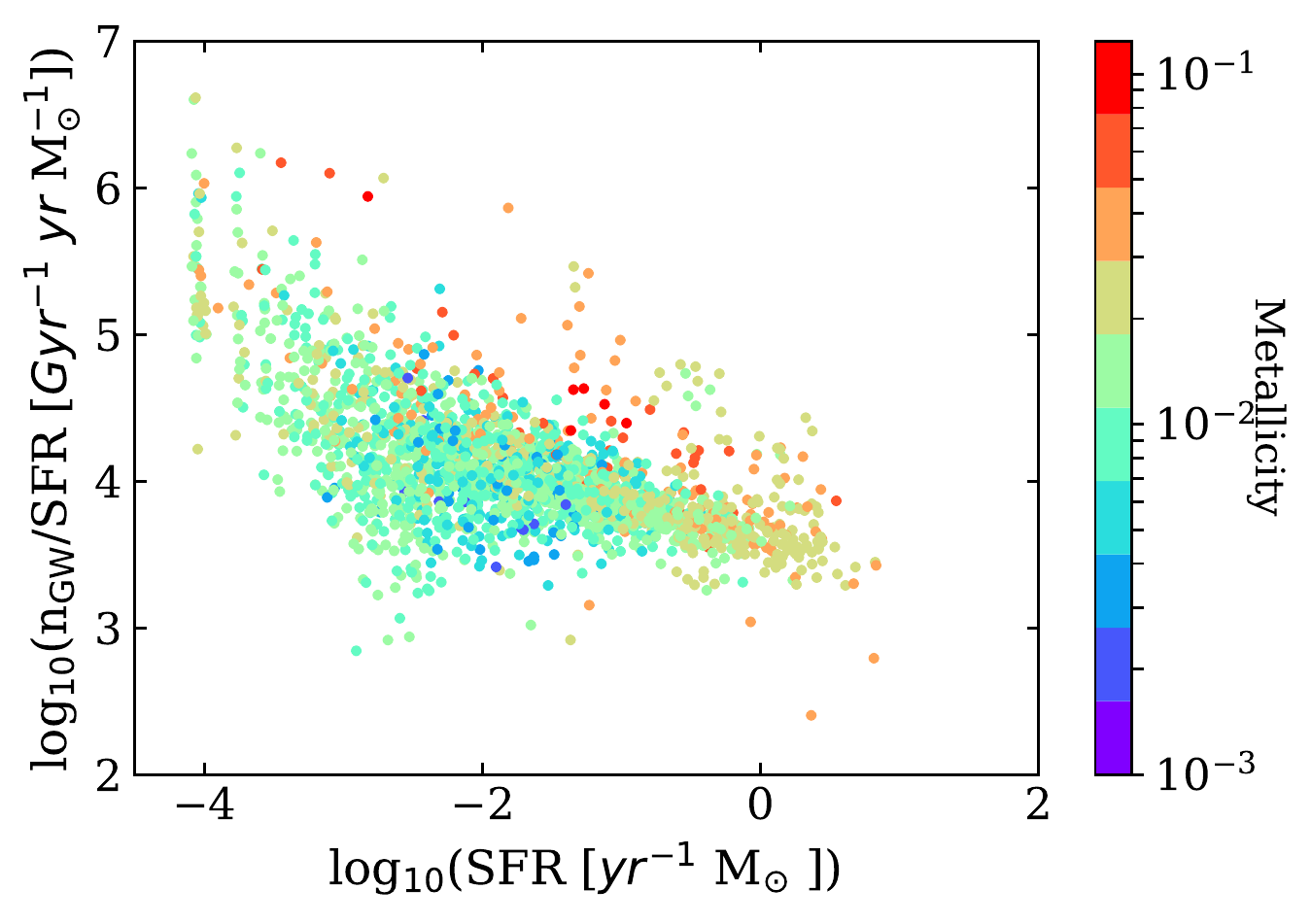}
\includegraphics[width=0.45\textwidth]{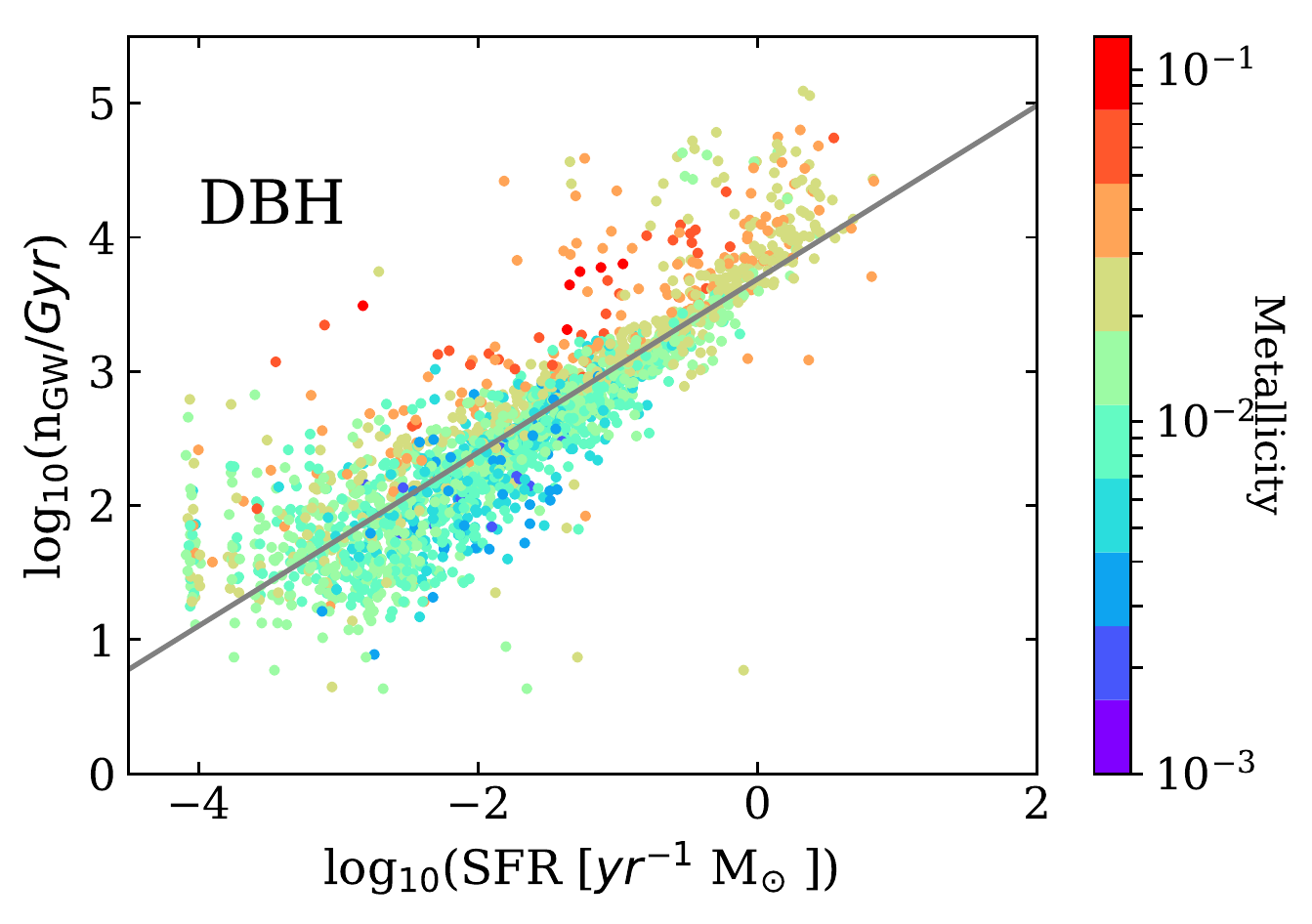}
\includegraphics[width=0.45\textwidth]{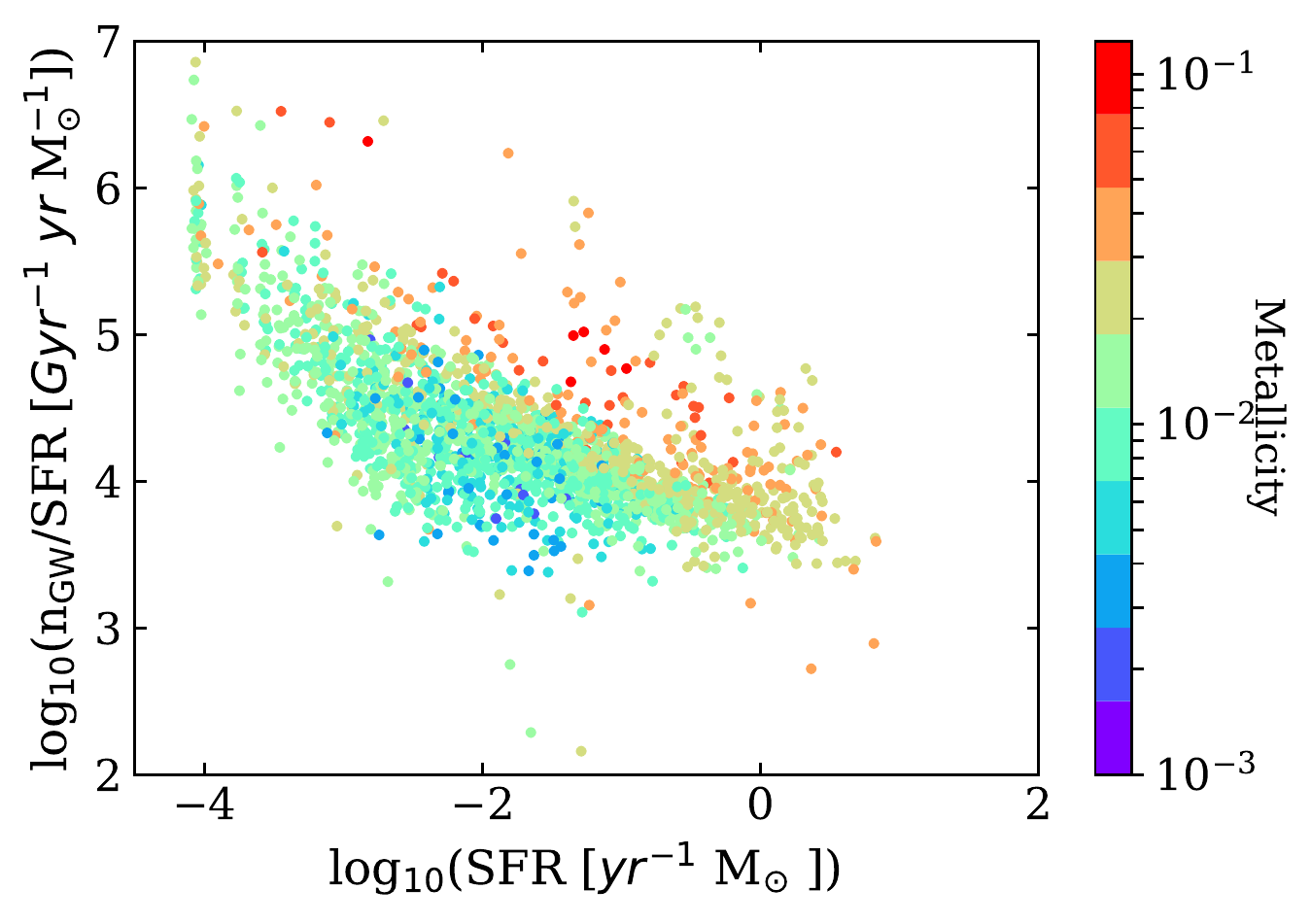}
\caption{\textit{Left-hand panel:} Merger rate per galaxy as a function of the star formation rate (SFR) for DNSs, BHNSs and DBHs. We include the fits of $\log ({\rm n_{GW}}/{\rm Gyr}) = a_{\rm SFR} \log(SFR) + b_{\rm SFR}$.
\textit{Right-hand panel:} Merger rate per galaxy normalized to the SFR, as a function of the SFR of the host galaxy. 
Each point represents an individual galaxy from the {\sc eagle} catalog.
The colour code represents the metallicity of the host galaxy.}
\label{fig:GW25Mpc-GWSFR}
\end{figure*}
%%%%%%%%%%%%%%%%%%%%%%%%%%%%%%%%%%%%%%%%%%%%%%%%%%%%%%%%%%%%%%%%%%%%%%%%%%%%%%

%%%%%%%%%%%%%%%%%%%%%%%%%%%%%%%%%%%%%FIGURE 5%%%%%%%%%%%%%%%%%%%%%%%%%%%%%%%%

\begin{figure}
\centering
\includegraphics[width=0.5\textwidth]{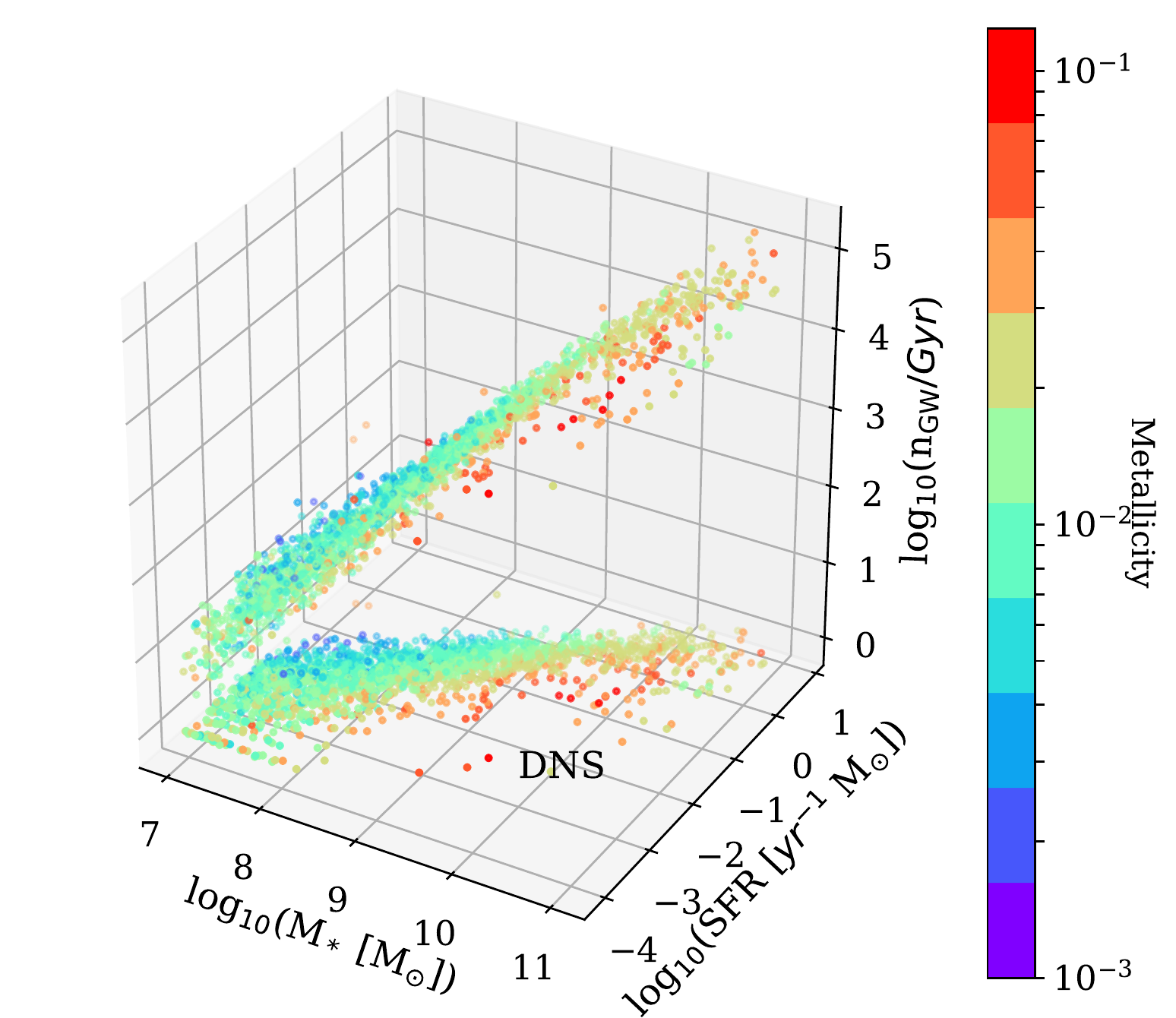}
\caption{3D map of the merger rate per galaxy ($z-$axis) as a function of SFR ($x-$axis) and stellar mass ($y-$axis) for the host galaxies of DNSs.
Each point represents an individual galaxy from the {\sc eagle} catalog.
The colour-code represents the metallicity of the galaxy. The points are also projected in the SFR -- M$_\ast$ relation.
}
\label{fig:DNS-fundamentalPlane}
\end{figure}
%%%%%%%%%%%%%%%%%%%%%%%%%%%%%%%%%%%%%FIGURE 5%%%%%%%%%%%%%%%%%%%%%%%%%%%%%%%%

The left-hand panel of Figure~\ref{fig:GW25Mpc-GWMs} shows the merger rate per galaxy (${\rm n}_{\rm GW}$) as a function of the stellar mass of the host galaxy.  
We find a very strong correlation between host galaxy mass and merger rate. This relation is steeper for DNSs than for DBHs and BHNSs, indicating that the dependence on stellar mass is more prominent for DNSs.
To quantify this difference, we fit the relation $\log{({\rm n_{GW}})} = a_{\rm M_\ast{}} \log(M_\ast{}) + b_{\rm M_\ast{}}$ with a least square linear regression\footnote{We use
the module polyfit from numpy, which minimises the squared error $E = \sum_{j=0}^{k}|p(x,k) - y_j|^2$ where $p(x,k)$ represents the linear function adopted, 
and $y_j$ refers to the data points \citep[see][]{NumpyBook,Numpy}. The reported errors are standard deviations computed using the diagonal of the covariance matrix for each parameter.}.

Table~\ref{tab:fits} shows the fit results. The correlation of the DNS merger rate per galaxy with the mass of the host galaxy is steeper than linear for DNSs, while it is sensibly shallower than linear for BHNSs and DBHs. % shows the parameters obtained from the linear fits for DNS, DBH and BHNS.

We find that galaxies with stellar mass $M_\ast = 10^{11}~\Msun$ host $\sim 4.4$ and $\sim 5.2$ more DNS mergers per Gyr than DBH and BHNS mergers, respectively.
 For galaxies with mass $M_\ast = 10^{9}~\Msun$ the difference is lower, reaching a factor of $\sim 1.9$. Our results indicate that massive galaxies are the best place to look for merging DNSs.  This result is in agreement with \citet{Mapelli2018}, which finds that the host galaxies of merging DNSs have preferentially stellar mass $>10^{9}~\Msun$.

For a fixed stellar mass, metal poor galaxies have a higher DBH and BHNS merger rate per galaxy than metal rich ones. This is apparent from the right-hand panel of Figure~\ref{fig:GW25Mpc-GWMs}, also shown in Figure~\ref{fig:MZR-GW25Mpc}. This dependence seems to be less significant for the host galaxies of DNSs.  
From the right panel of Figure~\ref{fig:GW25Mpc-GWMs} we find that low mass galaxies have a more efficient merger rate per galaxy of DBHs and BHNS systems, while for DNSs this trend is less significant.  

\subsection{Star formation rate}~\label{sec:Results_SFR}

 Figure~\ref{fig:GW25Mpc-GWSFR} shows the merger rate per galaxy (${\rm n}_{\rm GW}$) as a function of the SFR of the host galaxy. We find that the merger rate per galaxy correlates also with the SFR. 

 We fit the relation between the merger rate per galaxy and the SFR of the host galaxy using the same methodology previously described. Hence, we adopt the relation $\log ({\rm n_{GW}}) = a_{\rm SFR} \log({\rm SFR}) + b_{\rm SFR}$ (see Table~\ref{tab:fits}). Also in this case, the correlation is steeper for DNSs than for DBHs and BHNSs.
Moreover, the correlation between merger rate per galaxy and SFR is significantly less steep than the correlation between merger rate per galaxy and stellar mass. The former correlation has also a larger scatter than the latter one.

 It is interesting to note that for a fixed SFR, metal-rich galaxies are associated with a higher merger rate per galaxy than metal-poor ones (see Figure~\ref{fig:GW25Mpc-GWSFR}).
  However, this result must not be misunderstood, since selecting galaxies with a fixed SFR means putting together a wide range 
 of stellar masses. As stated by \citet{Mannucci2010}, galaxies in the local Universe show a fundamental relation between 
 the stellar mass, the gas-phase metallicity and the SFR. At a fixed stellar mass, galaxies with high SFR  have a lower metallicity than low SFR galaxies. 
  Moreover, at fixed SFR, massive galaxies are more metal rich compared with low mass galaxies.
 To clarify this and to make more explicit the connection with the fundamental metallicity relation, Figure~\ref{fig:DNS-fundamentalPlane} shows the relation between SFR, stellar mass and merger rate per galaxy for the host galaxies of DNSs.  
 Overall, galaxies with a higher SFR tend to be more massive than galaxies with a low SFR, as shown by \citet{Mannucci2010}.
 Hence, the correlation between merger rate per galaxy and SFR is likely a consequence of the correlation between SFR and stellar mass of the host galaxy.
 Therefore, our results reflect the strong correlation between the merger rate per galaxy and the stellar mass of the host galaxy.

%%%%%%%%%%%%%%%%%%%%%%%%%%%%%%%%%%%%%TABLE 1%%%%%%%%%%%%%%%%%%%%%%%%%%%%%%%%
\begin{table*}
\centering
\caption{Results of the linear fit of the compact-object (CO) merger rate per galaxy  ($n_{\rm GW}$) as a function of the stellar mass ($M_\ast$), the SFR and the metallicity.}
\label{tab:fits}
\begin{tabular}{|l||c|c|c|c|c|c|}
\hline
Merging COs          &   $a_{\rm M_\ast}$   &   $b_{\rm M_\ast}$   &   a$_{\rm SFR}$      &   b$_{\rm SFR}$   &   a$_{\rm Z}$      &   b$_{\rm Z}$    \\
\hline
DNSs                 &  1.15$\pm$0.08    &  -7.22$\pm$0.22   &  0.90$\pm$0.10       &   4.02$\pm$0.14        &   2.01$\pm$0.27    &  6.23$\pm$0.37  \\
DBHs                 &  0.80$\pm$0.07    &  -4.14$\pm$0.19   &  0.65$\pm$0.09       &   3.69$\pm$0.13        &   1.29$\pm$0.24    & 
4.99$\pm$0.33  \\
BHNSs                &  0.87$\pm$0.08    &  -4.99$\pm$0.22   &  0.70$\pm$0.09       &   3.56$\pm$0.13        &   1.09$\pm$0.26    &
4.38$\pm$0.35 \\
\hline
\end{tabular}
\end{table*}
%%%%%%%%%%%%%%%%%%%%%%%%%%%%%%%%%%%%%TABLE 1%%%%%%%%%%%%%%%%%%%%%%%%%%%%%%%%

\subsection{Host galaxy metallicity}
%%%%%%%%%%%%%%%%%%%%%%%%%%%%%%%%%%%%%FIGURE 6%%%%%%%%%%%%%%%%%%%%%%%%%%%%%%%%
\begin{figure}
\centering
\includegraphics[width=0.4\textwidth]{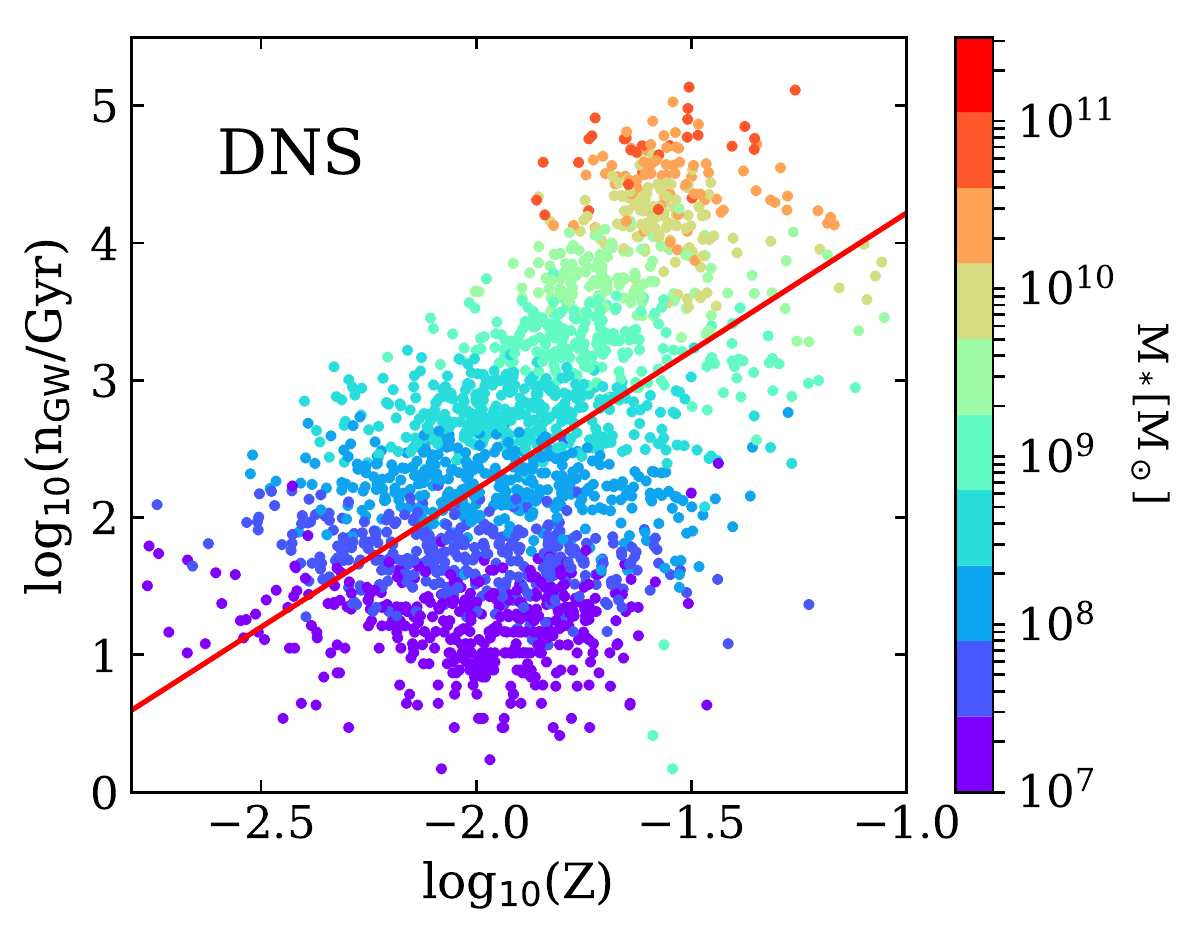}
\includegraphics[width=0.4\textwidth]{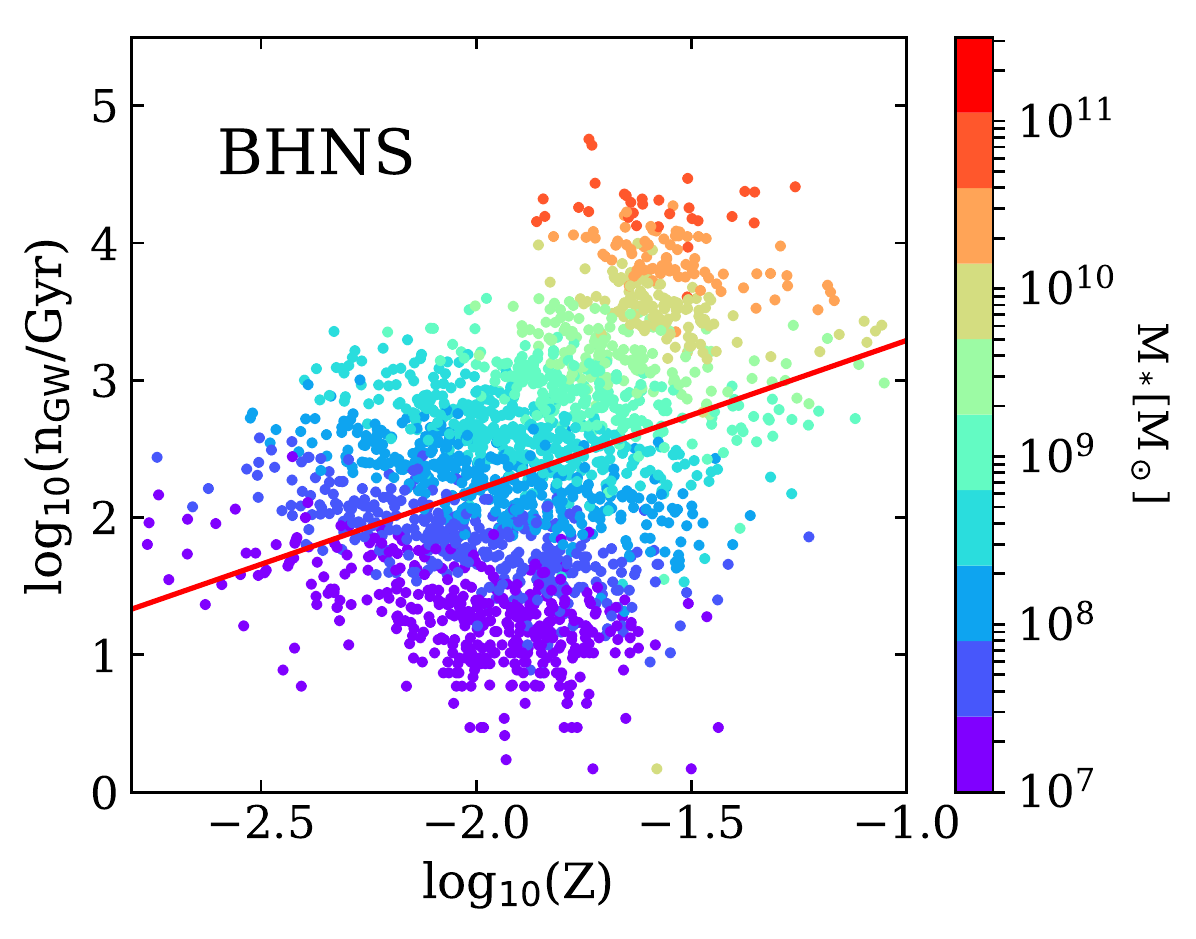}
\includegraphics[width=0.4\textwidth]{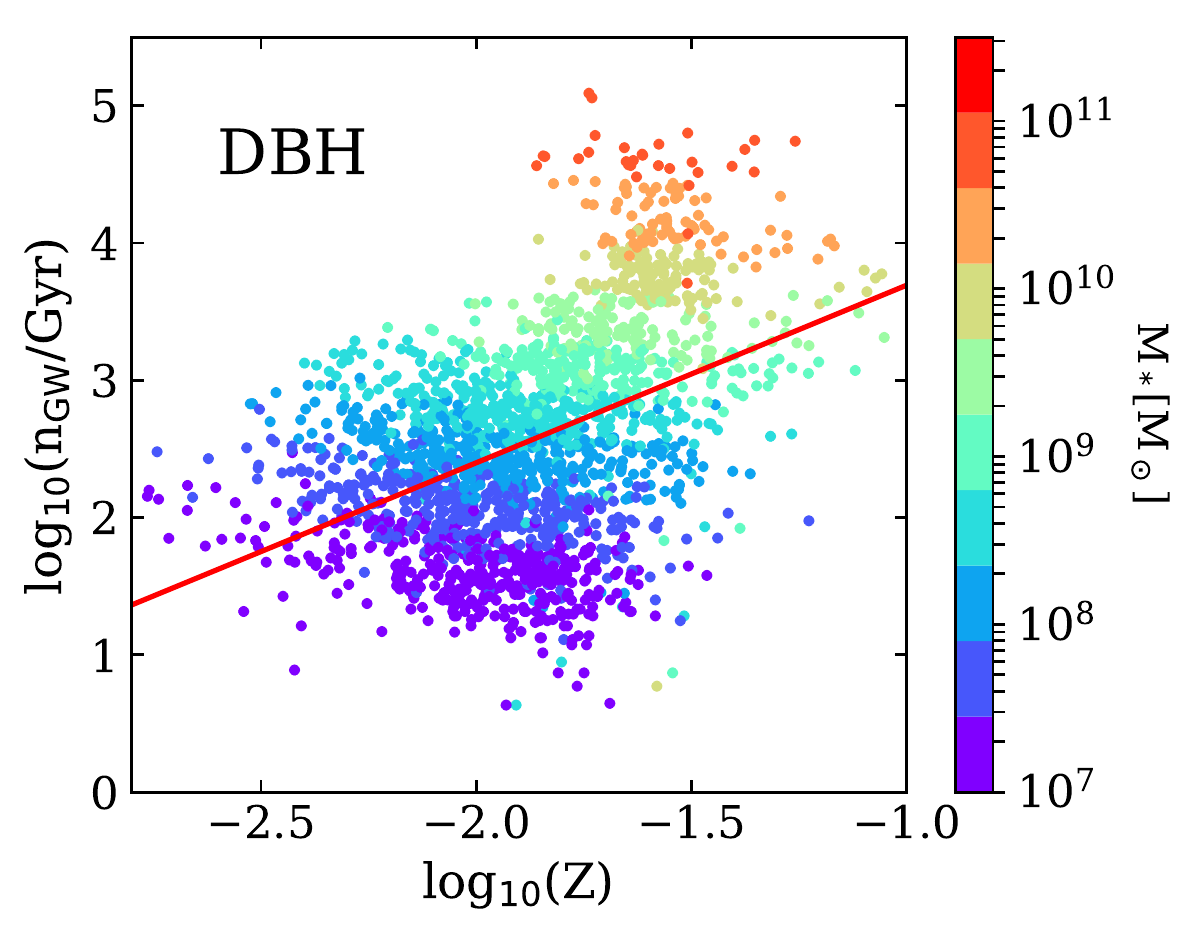}
\caption{Merger rate per galaxy as a function of the metallicity of the host galaxy at the time the compact objects merge.
The red lines represent the linear regression fits. Each point represents an individual galaxy from the {\sc eagle} catalog. The colour code represents the stellar mass of the galaxy.}
\label{fig:Z-NGW}
\end{figure}
%%%%%%%%%%%%%%%%%%%%%%%%%%%%%%%%%%%%%FIGURE 6%%%%%%%%%%%%%%%%%%%%%%%%%%%%%%%%

In Figure~\ref{fig:Z-NGW}, we show the merger rate per galaxy as a function of the metallicity for the host galaxies
of merging DNSs, DBHs, and BHNSs. We also include the linear regression fits computed as  $\log ({\rm n_{GW}}) = a_{\rm Z} \log(Z) + b_{\rm Z}$ (see the parameters obtained in Table~\ref{tab:fits}).
Metal rich galaxies have a higher merger rate since they also represent the most massive ones. This result holds not only for DNSs (which are almost not affected by progenitor's metallicity), but also for DBHs and BHNSs, which form predominantly from metal-poor progenitors. This indicates that the metallicity of the galaxy where these binaries merge is very different from the metallicity of the galaxy where these binaries formed. Overall, there is a very large scatter in the correlation between $Z$ and ${\rm n}_{\rm GW}$.

\subsection{The colour-magnitude diagram of the host galaxies of DNSs}

It is well known that the colour is a good tracer of SFR in galaxies, while luminosity correlates with their stellar mass.
Hence, exploring the colour--magnitude of the host galaxies can provide us information about where it is more likely to detect a GW event.
In this section we focus only on the host galaxies where DNSs merge, since only the host galaxy of GW170817 has been identified.

\begin{figure}
\centering
\includegraphics[width=0.5\textwidth]{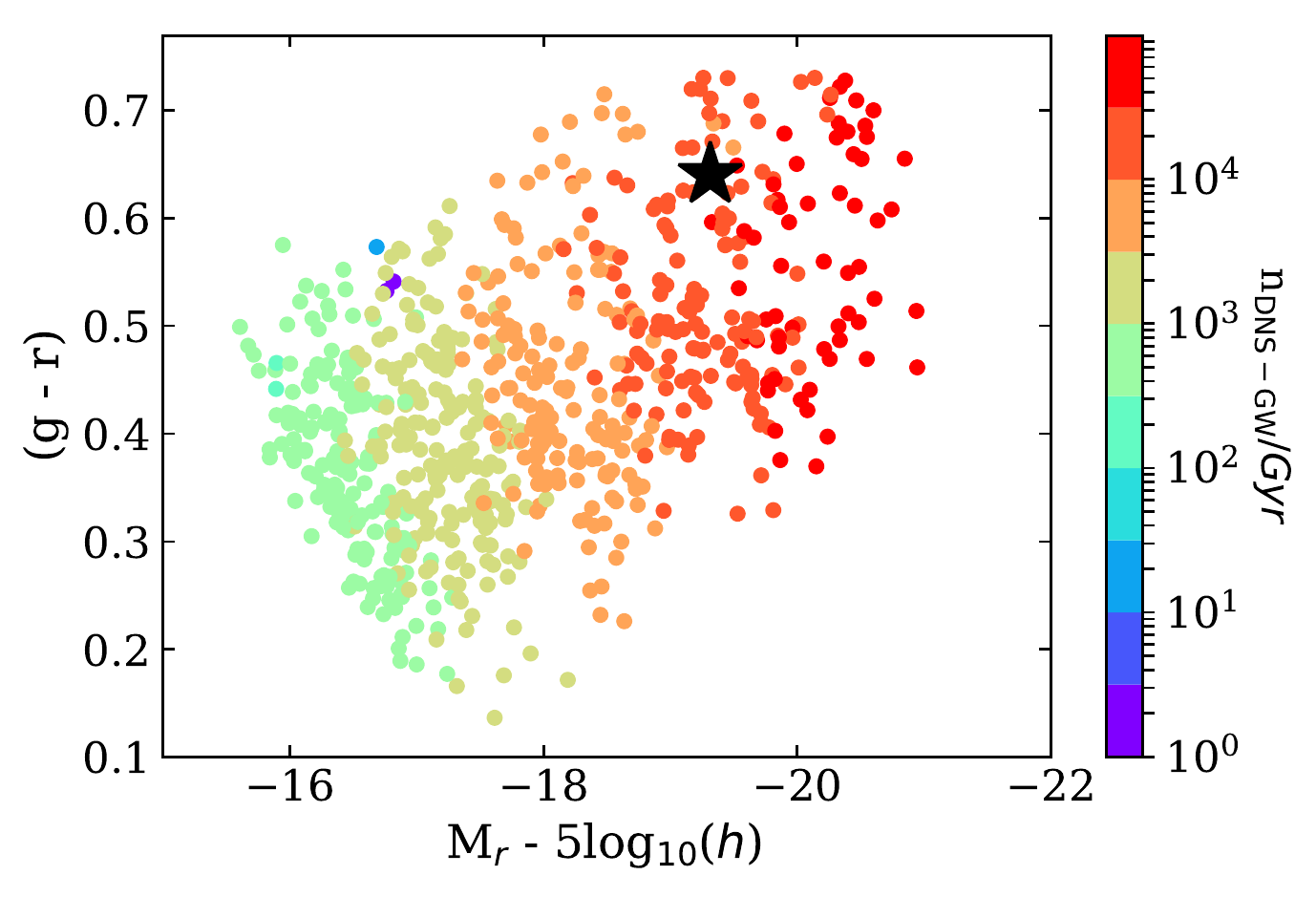}
\caption{Colour--magnitude diagram of the host galaxies of merging DNSs at $z\leq{}0.1$. Each point represents an individual galaxy from the {\sc eagle}. The colour code represents the number of merging DNSs per Gyr in each galaxy. The black star represents the location of NGC~4993, the host 
galaxy of GW170817 \citep[computed by][]{Blanchard2017} in the colour-magnitude diagram. 
} 
\label{fig:GW25MpcGalaxyColours}
\end{figure}

Figure~\ref{fig:GW25MpcGalaxyColours} shows the colour--magnitude diagram of the host galaxies of DNSs at their merger time, derived from the \eagle\ galaxy catalog. The sample of galaxies in this Figure is smaller than the sample considered in previous figures, because colours are only provided for galaxies with stellar masses above $10^{8.5}~\Msun$, containing more than 250 dust particles (see Section~\ref{sec:method}).

We find that brighter (massive) galaxies host a higher number of merging DNSs per Gyr, irrespective of their colour.
Our results confirm that the stellar mass is a much more important tracer of the merger rate than the SFR.
We also include the colour--magnitude of NGC~4993, the host galaxy of GW170817 computed by \citet{Blanchard2017}.
The reported magnitudes were converted to absolute and K-corrected to make a proper
comparison with \eagle\ galaxies.

From Figure~\ref{fig:GW25MpcGalaxyColours} it is apparent that NGC~4993 falls in the region of the colour-magnitude diagram associated with the higher merger rate per galaxy in the local Universe. This suggests that NGC~4993 is also one of the galaxies where it is more likely to detect DNS mergers, although to quantitatively support this statement we need to include observational selection effects and to convolve them with the instrumental range of GW detectors (Artale et al., in preparation).

%%%%%%%%%%%%%%%%%%%%%%%%%%%%%%%%%%%%%FIGURE 7%%%%%%%%%%%%%%%%%%%%%%%%%%%%%%%% 
%%%%%%%%%%%%%%%%%%%%%%%%%%%%%%%%%%%%%FIGURE 7%%%%%%%%%%%%%%%%%%%%%%%%%%%%%%%%

\section{Discussion}\label{sec:discussion}

We find a very strong correlation between the merger rate per galaxy ${\rm n}_{\rm GW}$ and the stellar mass of the host galaxy $M_\ast{}$ where the two compact objects merge. We find also a significant correlation between ${\rm n}_{\rm GW}$ and the SFR, and a very mild correlation between  ${\rm n}_{\rm GW}$ and the host galaxy metallicity.

It is worth noticing that the host galaxy where the two compact objects merge is not necessarily the same as the host galaxy where their progenitor binary star formed, because the initial host galaxy might have merged into a larger galaxy before the two compact objects reached coalescence \citep{Mapelli2018}. Moreover, even if the galaxy where the compact object merger occurs is the same as the galaxy where the progenitor binary formed, its SFR, stellar mass and metallicity at the time of the merger might be significantly different from the SFR, stellar mass and metallicity at the time of the formation of the stellar progenitors.

Thus, even if our model enforces a strong correlation between the merger rate of compact objects and the SFR at the time of the
progenitor binary formation, this does not imply that there is still a correlation at the time of the
merger. Thus, it is particularly remarkable that we still find a clear correlation between ${\rm n}_{\rm GW}$ and SFR. This correlation is likely 
a consequence of the strong correlation between SFR and galaxy stellar mass \citep{Mannucci2010}.

Similarly, even if our model enforces a strong anti-correlation between the metallicity of the progenitor star and the merger rate  of DBHs and BHNSs (not DNSs!), this does not imply that this anti-correlation still holds at the time of merger.  Indeed, we find a correlation (instead of an anti-correlation) between ${\rm n}_{\rm GW}$ and $Z$ in the case of DBHs and BHNSs. As we already discussed, the latter correlation is a consequence of the mass-metallicity relation (see Figures~\ref{fig:MZR-GW25Mpc} and \ref{fig:Z-NGW}).

The tight correlation between host galaxy mass and merger rate per galaxy is even more remarkable, because our model does not explicitly assume any link between the total mass of the galaxy where the compact objects merge and the merger rate. 

\begin{figure}
  \begin{center}
\includegraphics[width=0.4\textwidth]{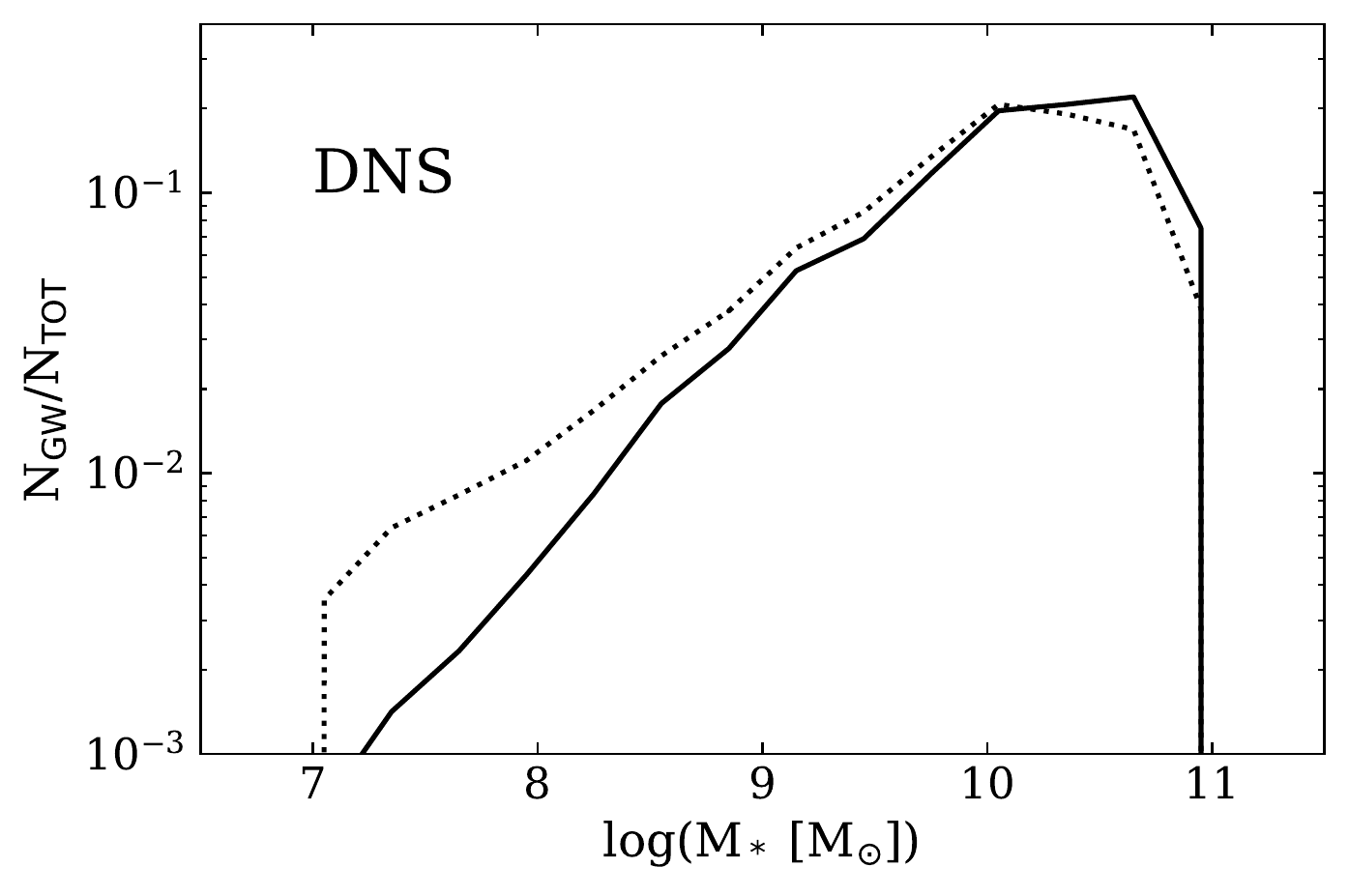} 
\includegraphics[width=0.4\textwidth]{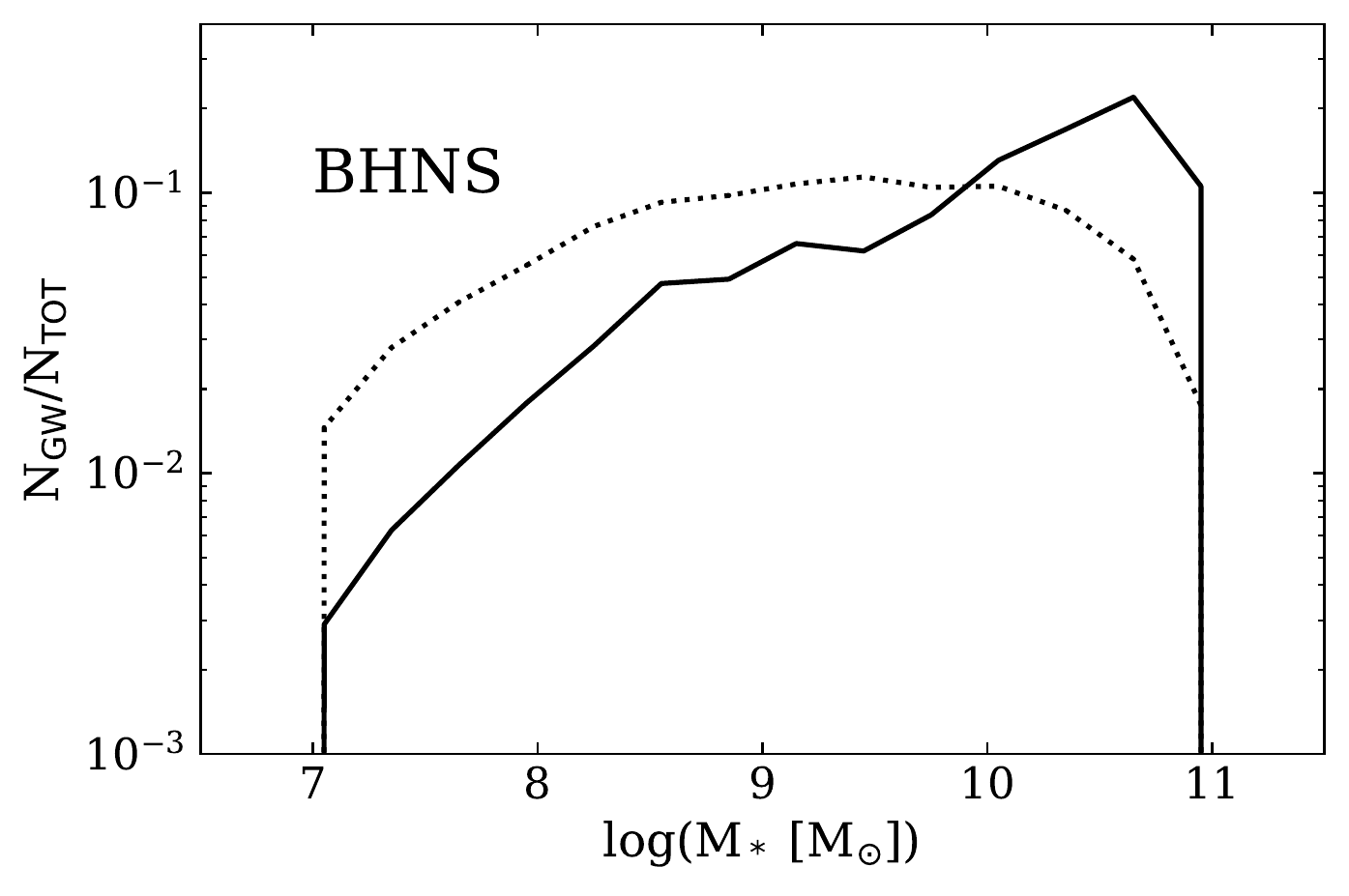} 
\includegraphics[width=0.4\textwidth]{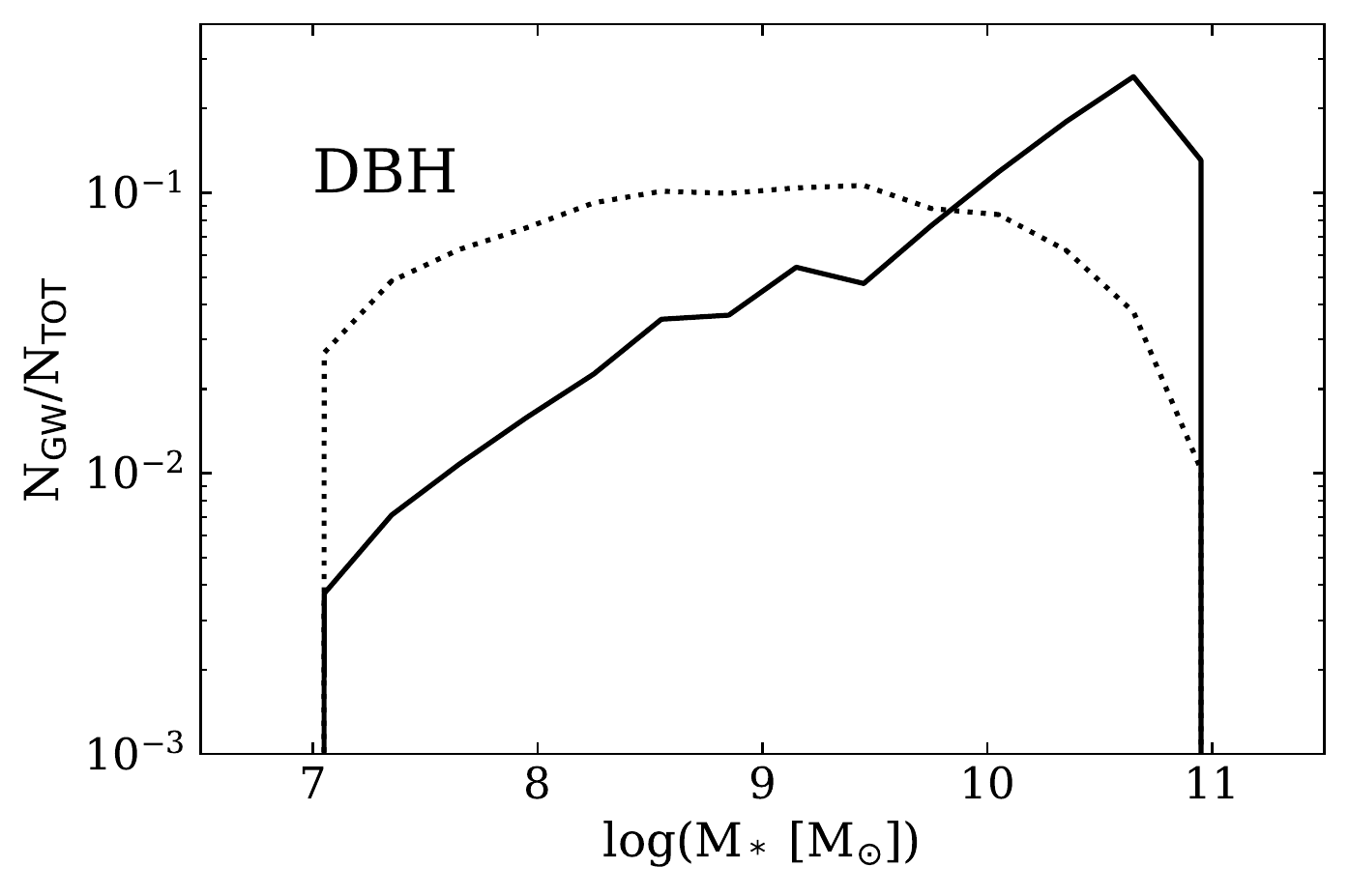} 
\caption{Stellar mass distribution of the host galaxies of merging DNSs (top), BHNSs (middle) and DBHs (bottom), where the binary systems formed (black dotted lines), and merge (black solid lines). The distribution is normalized by the total number of merging compact objects in each sample, N$_{\rm TOT}$.}
\label{fig:DistMasses_Form-Merge}
\end{center}
  \end{figure}

We find that the galaxies where DBHs, BHNSs and DNSs merge are mostly massive galaxies ($M_\ast{}\gtrsim{}10^{10}$ M$_\odot$). The galaxies where the progenitors of DNSs formed are also quite massive galaxies, while the galaxies where the progenitors of BHNSs and DBHs formed span a much larger mass range and tend to be skewed toward lower stellar masses (Figure~\ref{fig:DistMasses_Form-Merge}). This is in agreement with what already found by \cite{Mapelli2018}, and originates from the fact that most DBHs and BHNSs merging at $z\leq{}0.1$ form from metal-poor progenitors at high redshift and merge with long time delays, while  most DNSs merging at $z\leq{}0.1$ form from metal-rich progenitors and merge with short time delays.
\citet{Mapelli2018} combine the cosmological simulation Illustris-1 (which has lower resolution and adopts different sub-grid 
models with respect to the {\sc eagle}) with the population synthesis simulation CC15$\alpha{}$5 from {\sc mobse}. Hence, 
the good agreement between our findings and \citet{Mapelli2018} indicates that different resolution and different sub-grid physics in 
cosmological simulations do not affect our results significantly.

%%%%%%%%%%%%%%%%%%%%%%%%%%%%%%%%%%%%%FIGURE 8%%%%%%%%%%%%%%%%%%%%%%%%%%%%%%%%
%%%%%%%%%%%%%%%%%%%%%%%%%%%%%%%%%%%%%FIGURE 8%%%%%%%%%%%%%%%%%%%%%%%%%%%%%%%%

\begin{figure}
\centering
\includegraphics[width=0.45\textwidth]{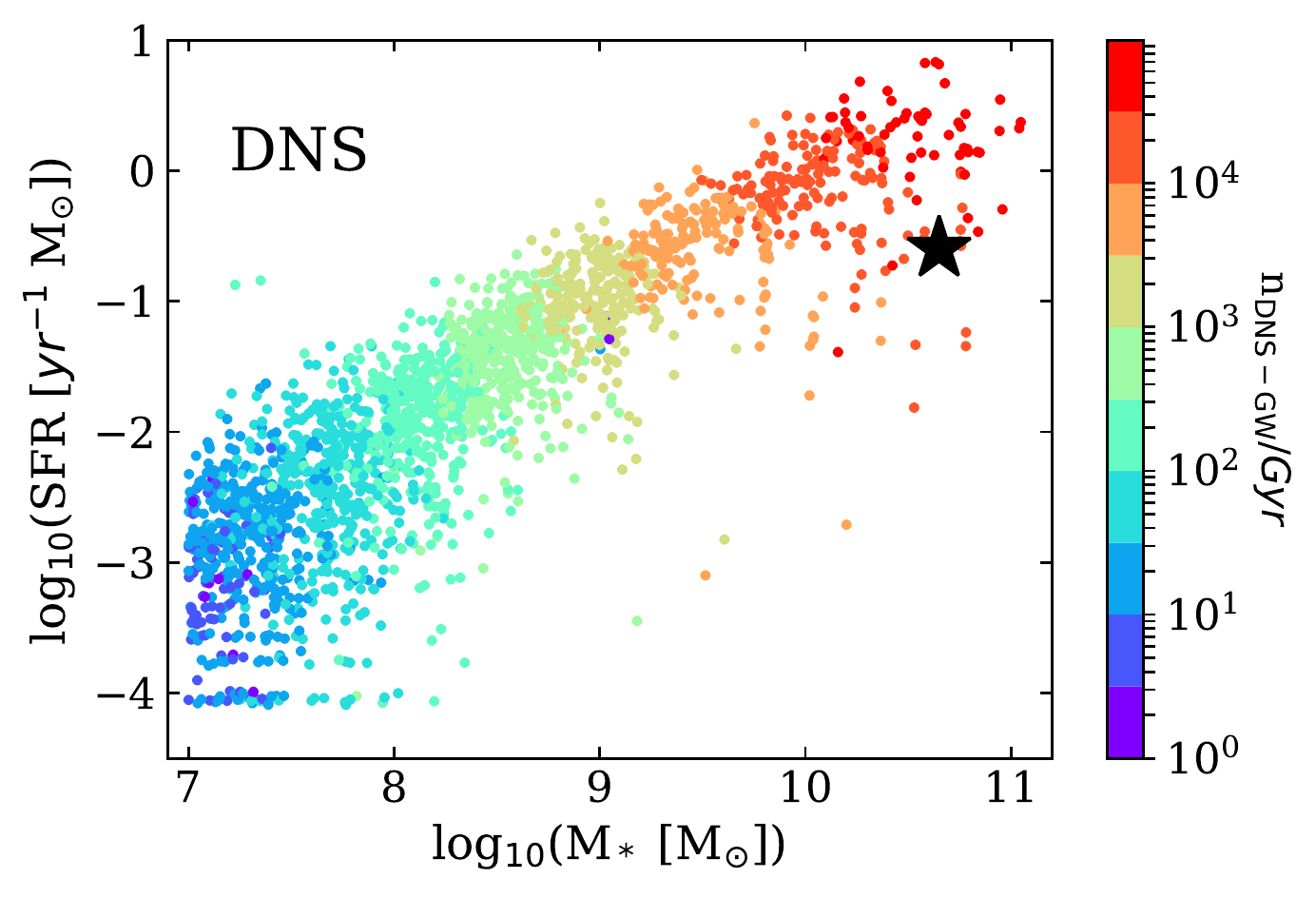}
\includegraphics[width=0.45\textwidth]{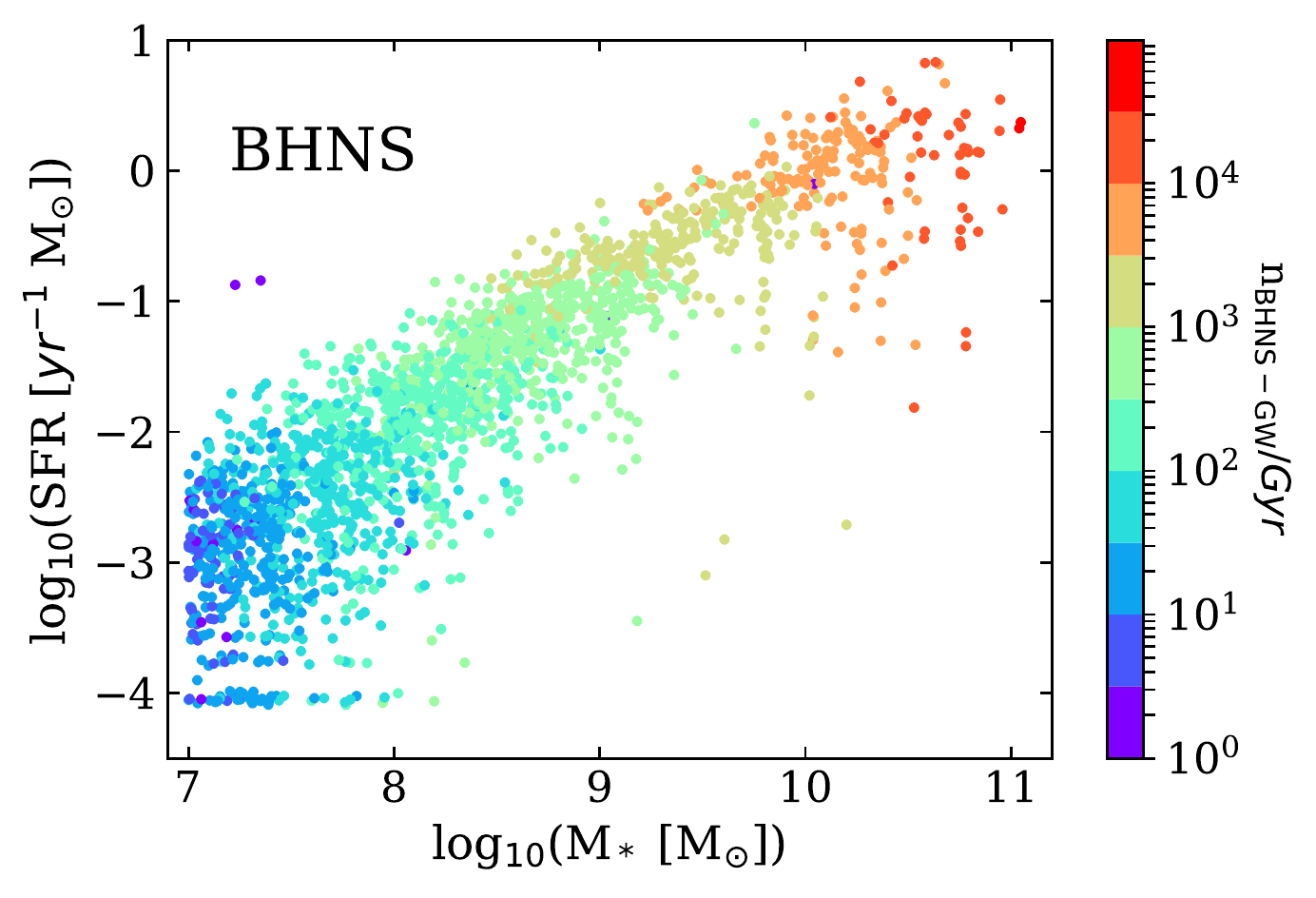}
\includegraphics[width=0.45\textwidth]{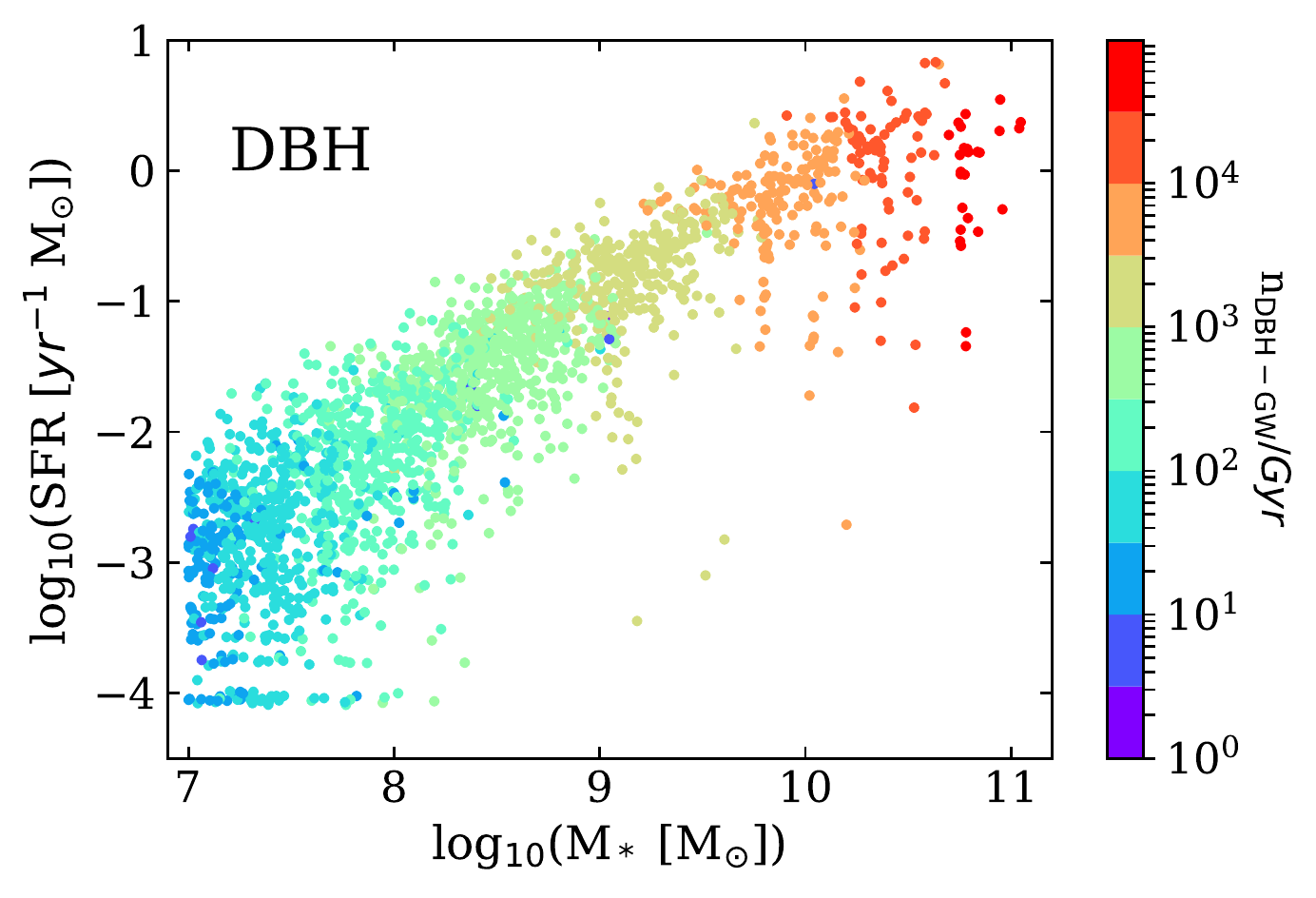}
\caption{SFR as a function of stellar mass for the host galaxies of merging DNSs, DBHs and BHNSs.
Each point represents an individual galaxy from the {\sc eagle} catalog.
The colour code represents the merger rate per each galaxy. 
The average SFR of NGC~4993 in the last Gyr (from $z\sim{}0.1$ to $z=0$) and its current stellar mass are indicated by the black star.
}
\label{fig:MsSFR-NGW}
\end{figure}

 In Figure~\ref{fig:MsSFR-NGW}, we plot the SFR as a function of the stellar mass of the host galaxies of DNSs, BHNSs and DBHs. At fixed stellar mass, the host galaxies present a wide range of SFRs, while the merger rate per galaxy remains roughly the same. This reaffirms the stronger dependence of the compact object merger rate on the stellar mass with respect to the SFR.

Table~\ref{tab:cumMergRate} shows the average merger rate per galaxy as the sum over all the merger rates per galaxy divided by the total number of galaxies for different sub-samples of galaxies ($r_{\rm GW}$). The average merger rate $r_{\rm GW}$ is maximum for galaxies with both large mass ($M_\ast{} \geq{}10^{10}$ M$_\odot$) and high SFR (SFR $\geq{}0.1$ M$_\odot$ yr$^{-1}$), but $r_{\rm GW}$ is still very large even for galaxies with large mass ($M_\ast{}\geq{}10^{10}$ M$_\odot$) and low SFR (SFR $<0.1$ M$_\odot$ yr$^{-1}$) due to the strong dependence with stellar mass presented in Section~\ref{sec:Results_Ms} and \ref{sec:Results_SFR}.

In contrast, the average specific merger rate ($r^{\rm spec}_{\rm GW}$, i.e. the average merger rate per unit stellar mass) shown in Table~\ref{tab:specMergRate} is larger for small star forming galaxies, consistently with what already found in previous work \citep{OShaughnessy2017}. This means that small galaxies with high star-formation rate are more efficient in producing merging binaries.

%%%%%%%%%%%%%%%%%%%%%%%%%%%%%%%%%%%%%FIGURE 9%%%%%%%%%%%%%%%%%%%%%%%%%%%%%%%% 
%%%%%%%%%%%%%%%%%%%%%%%%%%%%%%%%%%%%%FIGURE 9%%%%%%%%%%%%%%%%%%%%%%%%%%%%%%%%

%%%%%%%%%%%%%%%%%%%%%%%%%%%%%%%%%%%%%TABLE 2%%%%%%%%%%%%%%%%%%%%%%%%%%%%%%%%
\begin{table*}
\centering
\caption{Average merger rate per galaxy for the host galaxies of DNSs, BHNSs and DBHs at the time they merge. 
We split the galaxy sample by stellar mass, and SFR.} 
\label{tab:cumMergRate}
\begin{tabular}{|l||c|c|c|}
\hline
                                       &   DNS  &   BHNS   &   DBH    \\
\hline
r$_{\rm GW}(M_\ast{} \geq 10^{10}\Msun \&\,{} {\rm SFR} \geq 0.1\Msunyr)/\Gyr$  &  34640  &  9680    &  19120     \\
r$_{\rm GW}(M_\ast{}  \geq 10^{10}\Msun  \&\,{} {\rm SFR} < 0.1\Msunyr)/\Gyr$    &  9920  &  5770    &  14310     \\
r$_{\rm GW}(M_\ast{} < 10^{10}\Msun \&\,{} {\rm SFR} \geq 0.1\Msunyr)/\Gyr$     &  5740   &  1720    &  2430     \\
r$_{\rm GW}(M_\ast{} < 10^{10}\Msun \&\,{} {\rm SFR} < 0.1\Msunyr)/\Gyr$        &  210    &  150     &  240   \\
\hline
\end{tabular}
\end{table*}
%%%%%%%%%%%%%%%%%%%%%%%%%%%%%%%%%%%%%TABLE 2%%%%%%%%%%%%%%%%%%%%%%%%%%%%%%%%

%%%%%%%%%%%%%%%%%%%%%%%%%%%%%%%%%%%%%TABLE 3%%%%%%%%%%%%%%%%%%%%%%%%%%%%%%%%
 \begin{table*}
 \centering
\caption{Average specific merger rate r$^{\rm spec}_{\rm GW}$ for the host galaxies of DNSs, BHNSs and DBHs at the time they merge. We split the galaxy sample by stellar mass, SFR and sSFR.
 We refer as early-type galaxies to those galaxies with ${\rm sSFR} < 10^{-10} {\rm yr}^{-1}$, while late-type galaxies have a ${\rm sSFR} \geq 10^{-10} {\rm yr}^{-1}$.}
 \label{tab:specMergRate}
 \begin{tabular}{|l||c|c|c|}
 \hline
                                                        &   DNS                 &   BHNS                 &   DBH    \\
 \hline
r$^{\rm spec}_{\rm GW}(M_\ast{} \geq 10^{10}\Msun \&\,{} {\rm SFR} \geq 0.1\Msunyr)/10^{7}\Msun\Gyr$  & 5.9  &  1.7    &   3.4   \\
r$^{\rm spec}_{\rm GW}(M_\ast{} \geq 10^{10}\Msun  \&\,{} {\rm SFR} < 0.1\Msunyr)/10^{7}\Msun\Gyr$    & 2.2  & 1.3     &  3.3     \\
r$^{\rm spec}_{\rm GW}(M_\ast{} < 10^{10}\Msun \&\,{} {\rm SFR} \geq 0.1\Msunyr)/10^{7}\Msun\Gyr$     & 10.1  &  3.1    &  4.4     \\
r$^{\rm spec}_{\rm GW}(M_\ast{} < 10^{10}\Msun \&\,{} {\rm SFR} < 0.1\Msunyr)/10^{7}\Msun\Gyr$        & 5.8  &  4.3     &  7.0   \\
\hline
r$^{\rm spec}_{\rm GW}(\rm sSFR \geq{} 10^{-10} \yr^{-1})/10^{7}\Msun\Gyr$  & 11.4  &  3.5 & 4.7  \\
r$^{\rm spec}_{\rm GW}(\rm sSFR < 10^{-10} \yr^{-1})/10^{7}\Msun\Gyr$      & 5.0 & 1.7 & 3.6   \\
 \hline
 \end{tabular}
 \end{table*}
%%%%%%%%%%%%%%%%%%%%%%%%%%%%%%%%%%%%%TABLE 3%%%%%%%%%%%%%%%%%%%%%%%%%%%%%%%%

%%%%%%%%%%%%%%%%%%%%%%%%%%%%%%%%%%%%%TABLE 4%%%%%%%%%%%%%%%%%%%%%%%%%%%%%%%%
\begin{table}
\centering
\caption{Local merger rate density for early-type (R$^{\rm ET}$) and late-type (R$^{\rm LT}$) galaxies from the \eagle\ simulation.}
\label{tab:tablerates}
\begin{tabular}{|l||c|c|}
\hline
                                       &   $R_{\rm ET}$            &   $R_{\rm LT}$   \\
                                       &   $[\Gpc^{-3} \yr^{-1}]$  &   $[\Gpc^{-3} \yr^{-1}]$ \\
\hline
DNS                               &      146            &      92           \\
BHNS                             &        50              &    28           \\
DBH                               &      104            &      38          \\
\hline
\end{tabular}
\end{table}
%%%%%%%%%%%%%%%%%%%%%%%%%%%%%%%%%%%%%TABLE 4%%%%%%%%%%%%%%%%%%%%%%%%%%%%%%%%

\subsection{NGC~4993}
The fact that the correlation between merger rate per galaxy and stellar mass  is significantly steeper than the correlation between merger rate per galaxy and SFR also provides a valuable hint to understand GW170817.

It has been argued (e.g. \citealt{Chruslinska2018,Belczynski2019}) that finding the first DNS merger within an early-type galaxy might be in tension with models, because the SFR of NGC~4993 is low, while all models assume a correlation between the merger rate and the SFR. Our results show that the total stellar mass of the galaxy has more impact on the merger rate per galaxy than its current SFR, although our model assumed a strong correlation with the SFR at the time of progenitors' formation.

Indeed, by looking at Figure~\ref{fig:GW25MpcGalaxyColours} and at Figure~\ref{fig:MsSFR-NGW}, we find that the properties of NGC~4993 characterise it as one of the galaxies with the highest local merger rate per galaxy in our model.
Furthermore, as discussed previously, Table~\ref{tab:cumMergRate} shows that the average merger rate $r_{\rm GW}$ is still large for galaxies with large mass ($M_\ast{}\geq{}10^{10}$ M$_\odot$) and low SFR (SFR $<0.1$ M$_\odot$ yr$^{-1}$), similar to NGC~4993. In particular, we predict a current DNS merger rate $\mathcal{R}_{\rm NGC4993}\sim{}3-107$ Myr$^{-1}$ for a galaxy with the same stellar mass and SFR as NGC~4993.

We know that $\sim{}36$~\%  of the total stellar mass in the local Universe is locked inside elliptical galaxies (see Table~3 of \citealt{moffett2016}). This percentage rises to $\sim{}72$~\% if we consider all early type galaxies: not only ellipticals but also S0 and Sa galaxies (NGC~4993 is an S0 galaxy, \citealt{Levan2017}). Thus, most stellar mass in the local Universe is located in early-type galaxies such as NGC~4993. This observational information, combined with the relatively high merger rate that we estimated for NGC~4993, helps us understanding why the observation of the first DNS merger in an early-type galaxy is not surprising, but is  in agreement with current models.

\subsection{The Milky Way}

From our models, a galaxy with a typical stellar mass of $\sim{}5 \times{}10^{10}$ M$_\odot$ \citep{BlandHawthorn2016}
and with a current SFR of $\sim{}1.65$ M$_\odot$ yr$^{-1}$, i.e. a Milky-way like galaxy \citep{Licquia2015}, should have a typical DNS merger rate per galaxy of $\mathcal{R}_{\rm MW}\sim{}16-121$ Myr$^{-1}$, which is perfectly consistent with the Galactic DNS merger rate estimated by \cite{Pol2019} ($\mathcal{R}_{\rm MW}=42^{+30}_{-14}$ Myr$^{-1}$).

Thus, our main results are consistent with both the merger rate density of DNSs inferred from LIGO-Virgo and with the Galactic DNS merger rate.

\subsection{The local merger rate density from the \eagle{}: early-type versus late-type galaxies}
Finally, we can estimate the local merger rate density directly from the \eagle{}. Moreover, we can distinguish between the local merger rate density from late-type galaxies and from early-type galaxies, an information we cannot derive directly from GW detections. We stress that in the  \eagle\ simulation (25Mpc box), the total stellar mass locked up in early-type galaxies and late-type galaxies is $29\times10^{7} \Msun \Mpc^{-3}{}h_{0.7}$ (where $h_{0.7}=h/0.7$ is the Hubble parameter) and $8.1\times10^7 \Msun \Mpc^{-3}{}h_{0.7}$, respectively\footnote{Here, we define early-type galaxies (late-type galaxies) as galaxies with specific SFR $\leq{}10^{-10}$ yr$^{-1}$ ($>10^{-10}$ yr$^{-1}$).}. These values are in fair agreement with observations \citep{moffett2016}, from which we know that $\sim{}72$~\% of the total stellar mass in the local Universe is located in early-type galaxies.

From our results, we obtain a local DNS merger rate density $R_{\rm ET}\sim{}146$~Gpc$^{-3}$ yr$^{-1}$ $h_{0.7}$  and $R_{\rm LT}\sim{} 92$~Gpc$^{-3}$ yr$^{-1}$ $h_{0.7}$ from early-type and late-type galaxies, respectively (see Table~\ref{tab:tablerates}). Thus, the total local DNS merger rate density from the \eagle{} is $\sim{}238$~Gpc$^{-3}$ yr$^{-1}$ $h_{0.7}$, inside the LIGO-Virgo estimated rate ($\sim{}110-3840$ Gpc$^{-3}$ yr$^{-1}$, \citealt{AbbottO2a}).

With the same approach, we can estimate also the merger rate of DBHs and BHNSs (see Table~\ref{tab:tablerates}). About 73~\% (64~\%) of all DBH (BHNS) mergers we expect in the local Universe happen in early-type galaxies, confirming the same trend as DNSs. 
The total local DBH merger rate density we estimate from the \eagle{} is marginally too high if compared to the merger rate estimated from the LIGO-Virgo collaboration 
($\sim{}9.7-101$ Gpc$^{-3}$ yr$^{-1}$ for DBHs and an upper limit of $\sim{}610$ Gpc$^{-3}$ yr$^{-1}$ for BHNSs, \citealt{AbbottO2a}), but to compare these rates more properly we should estimate the detection rate we expect from our simulations. In fact, the merger rate estimated from the LIGO-Virgo collaboration assumes a mass spectrum of BHs which is sensibly different from the mass distribution we obtain from our population-synthesis simulations (the estimated merger rate of DNSs is less affected by this problem, because the mass range of DNSs is significantly smaller than the one of DBHs).

%%%%%%%%%%%%%%%%%%%%%%%%%%%%%%%%%%%%%FIGURE 10%%%%%%%%%%%%%%%%%%%%%%%%%%%%%%%%
\begin{figure}
  \begin{center}
\includegraphics[width=0.4\textwidth]{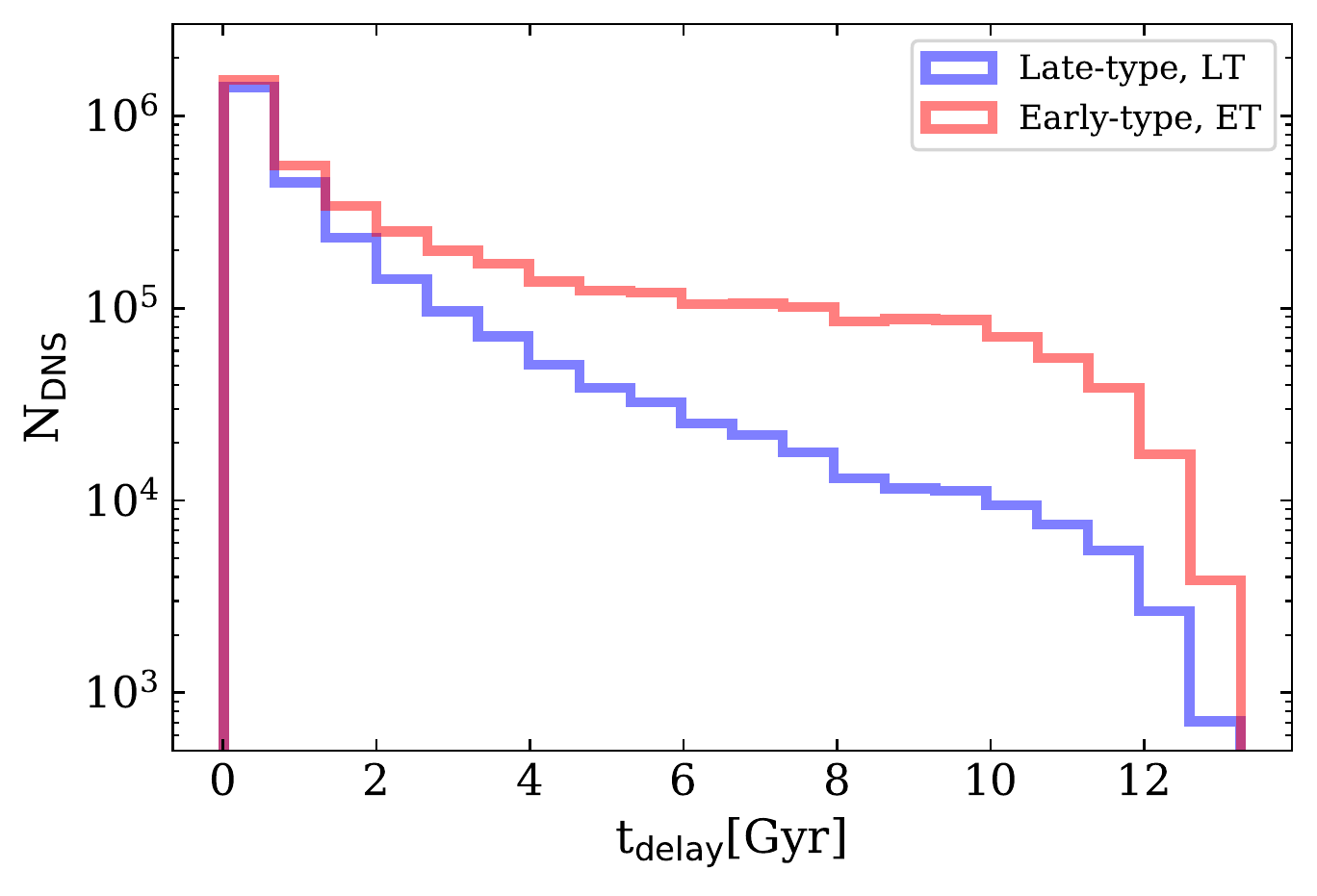} 
\includegraphics[width=0.4\textwidth]{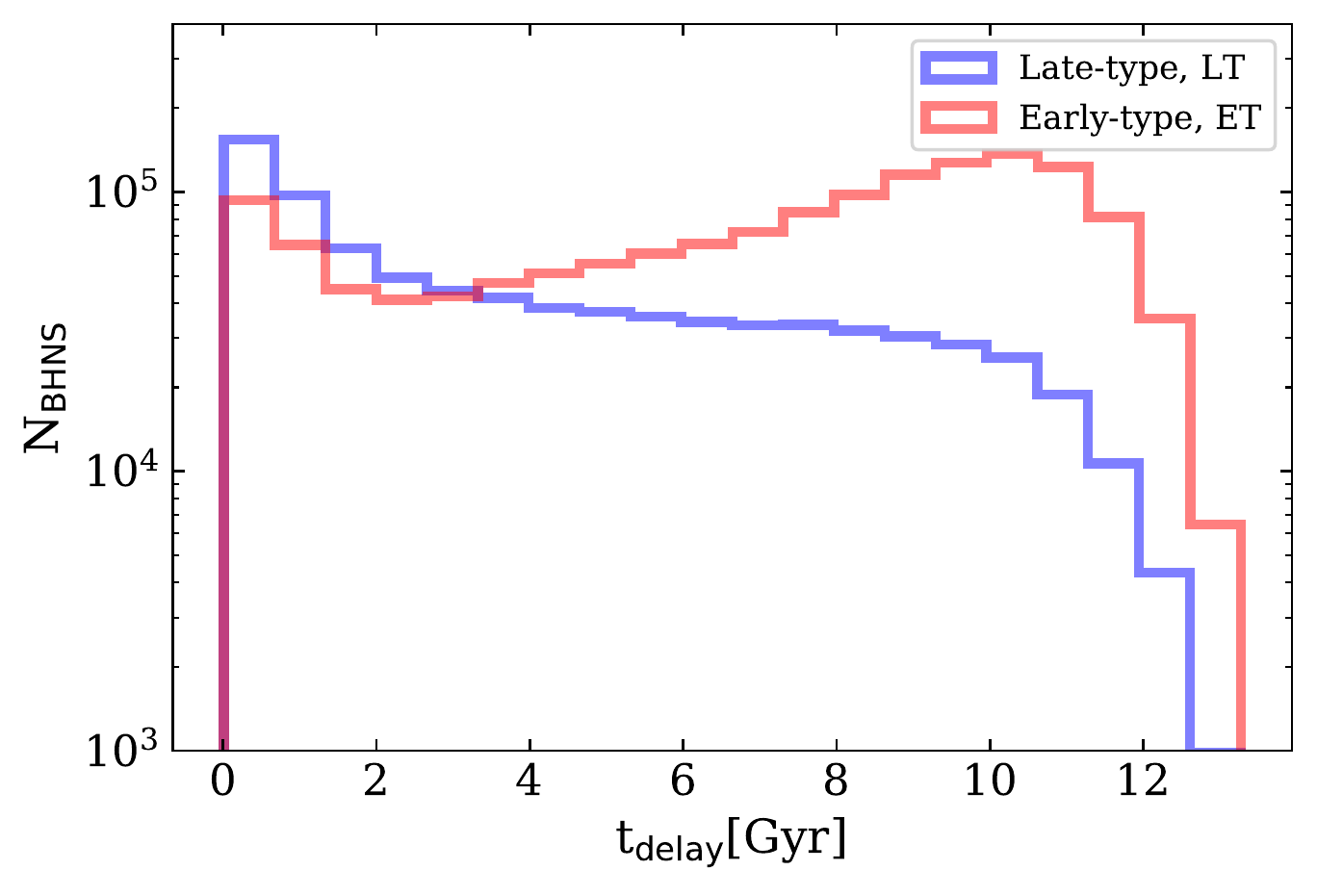} 
\includegraphics[width=0.4\textwidth]{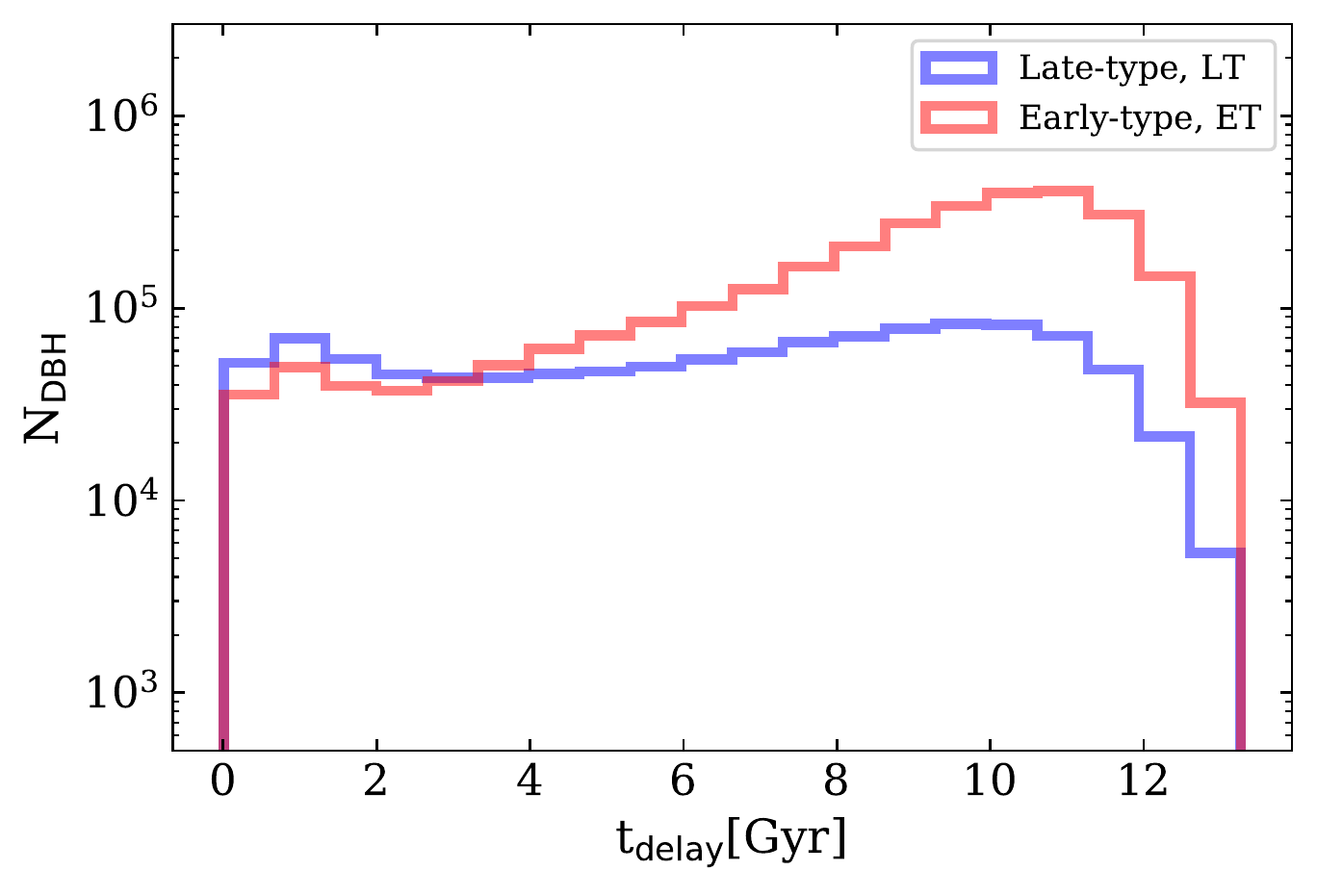} 
\caption{Distribution of delay times t:delay of DNSs (top), BHNSs (middle) and DBHs (bottom)
in late-type (blue histogram) and early-type (red histogram) galaxies in the local Universe.}
\label{fig:delayT}
\end{center}
  \end{figure}
%%%%%%%%%%%%%%%%%%%%%%%%%%%%%%%%%%%%%FIGURE 10%%%%%%%%%%%%%%%%%%%%%%%%%%%%%%%%
\subsection{Delay times}
As a corollary of our results, we expect that the distribution of delay times $t_{\rm delay}$ (i.e. the time elapsed between the formation of the progenitor stars and the compact-object merger) depends on the environment of the host galaxy.

If we consider a coeval stellar population (as we obtain directly from the {\sc MOBSE} simulations, before combining them with the cosmological simulation), the distribution of delay times scales approximately as $dN/dt\propto{}t_{\rm delay}^{-1}$ (see e.g. \citealt{Dominik2012,Giacobbo2018,Mapelli2018}).

In contrast, if we consider the delay times of all compact objects merging in a given time bin (regardless of their formation redshift), the distribution of delay times can be significantly different from  $dN/dt\propto{}t_{\rm delay}^{-1}$. In particular, Figure~4 of \cite{Mapelli2018} shows that the distribution of delay times of compact objects merging in the local Universe can be significantly flatter than $t_{\rm delay}^{-1}$, especially for DBHs and BHNSs. Similarly, \cite{mapelli2019} suggest that $>95$~\% of DBHs merging in the local Universe have $t_{\rm delay}>1$ Gyr (see their Table~2). 

Here, we investigate the dependence of $t_{\rm delay}$ on the environment, and in particular on the SFR of the host galaxy. Figure~\ref{fig:delayT} shows the delay time distribution of compact objects merging in the local Universe ($z\leq{}0.1$) in early-type galaxies and late-type galaxies, respectively. The two distributions are dramatically different, as expected. In particular, the delay time distribution in late-type galaxies is reminiscent of the  $t_{\rm delay}^{-1}$ scaling, because the population of merging systems is dominated by compact objects which formed recently. In contrast, the bulk of merging compact objects in early-type galaxies has a long delay time (peaked at $t_{\rm delay}\sim{}10$ Gyr), because the main episodes of star formation in these galaxies ended several Gyr ago. Thus,  most  compact objects merging in early-type galaxies in the local Universe formed several Gyr ago.

\section{Conclusions}\label{sec:conclusions}

We have investigated the properties of the host galaxies of double neutron stars (DNSs), double black holes (DBHs) and black hole-neutron star (BHNSs) at the time they merge in the local Universe ($z\leq{}0.1$), by combining the population-synthesis code {\sc mobse} \citep{Giacobbo2018} with galaxy catalogs from the \eagle\ simulation \citep{Schaye15}.

The results of {\sc mobse} are consistent with compact-object masses and merger rates reported by LIGO-Virgo detections \citep{Mapelli2017,Mapelli2018,Giacobbo2018B,Giacobbo2018c}.

In this work, we focus on the  stellar mass, star formation rate, metallicity and colours of the host galaxies of compact object mergers. These are fundamental properties of the host galaxies and can help us characterizing  the environment of merging compact objects. 

Our results show that the stellar mass $M_\ast$ of the host galaxy is an excellent tracer of the merger rate per galaxy ${\rm n}_{\rm GW}$ in the local Universe (Figure~\ref{fig:GW25Mpc-GWMs}). Massive galaxies have a higher merger rate with respect to low mass galaxies. The star formation rate of the host galaxies also correlates with the merger rate, but the SFR$-{\rm n}_{\rm GW}$ correlation is less tight than the $M_\ast{}-{\rm n}_{\rm GW}$ correlation (Figure~\ref{fig:GW25Mpc-GWSFR}).
The SFR$-{\rm n}_{\rm GW}$ correlation is likely a mere consequence of the fact that the SFR correlates with the stellar mass of the galaxy (Figure~\ref{fig:MsSFR-NGW}).

As a consequence, we also find a correlation of ${\rm n}_{\rm GW}$ with the  $g-r$ colour (which is a proxy of the SFR) and with the $r$ luminosity (which is a proxy of the stellar mass) of the host galaxy (Figure~\ref{fig:GW25MpcGalaxyColours}). 

Finally, we also find a loose correlation between ${{\rm n}_{\rm GW}}$ and the average metallicity $Z$ of the host galaxy at the time of merger (Figure~\ref{fig:Z-NGW}). This correlation is a by-product of the mass-metallicity relation of galaxies \citep{Maiolino2008}. Note that our population-synthesis models enforce an anti-correlation between progenitor's metallicity and the merger efficiency of DBHs and BHNSs.

These correlations might be crucial for the localization of  electromagnetic counterparts. Following-up on this, in a forthcoming work (Artale et al., in preparation), we will estimate the probability that a given galaxy hosts a compact-object merger as a function of its main properties.

We show that $\gtrsim{}60$~\% DNSs mergers in the local Universe are expected to happen in early-type galaxies ($R_{\rm ET}\sim{}146$~Gpc$^{-3}$ yr$^{-1}$ $h_{0.7}$), while only $\lesssim{}40$~\% DNSs mergers are expected in late-type galaxies ($R_{\rm LT}\sim{} 92$~Gpc$^{-3}$ yr$^{-1}$ $h_{0.7}$). This comes from the fact that $\sim{}78$~\% of the stellar mass in the \eagle{} box is locked up in early-type galaxies, in agreement with the observational results by \cite{moffett2016}. 

In particular, NGC~4993-like galaxies (i.e. galaxies with the same stellar mass and SFR of the host galaxy of GW170817) have a merger rate per galaxy $\mathcal{R}_{\rm NGC4993}\sim{}3-107$ yr$^{-1}$. Our results suggest that massive early-type galaxies like NGC~4993 are characterized by a relatively large merger rate per galaxy, due to their large mass.

Furthermore, we expect that Milky-Way like galaxies host a DNS merger rate $\mathcal{R}_{\rm MW}\sim{}16-121$ Myr$^{-1}$, which is consistent with the Galactic DNS merger rate estimated by \cite{Pol2019} ($\mathcal{R}_{\rm MW}=42^{+30}_{-14}$ Myr$^{-1}$). Thus, our results are in agreement with both the DNS merger rate derived from the LIGO-Virgo collaboration (based on GW170817) and the DNS merger rate estimated from Galactic DNSs \citep{Pol2019}.

\section*{Acknowledgement}
We thank the anonymous referee for their useful comments, and we are grateful to Stefania Marassi and Luca Graziani for their
suggestions.
MCA and MM acknowledge financial support from the Austrian National Science Foundation through FWF stand-alone grant P31154-N27
``Unraveling merging neutron stars and black hole -- neutron star binaries with population synthesis simulations''.
MM acknowledges financial support by the European Research Council for the ERC Consolidator grant DEMOBLACK, under contract no. 770017.
 MS acknowledges funding from the European Union's Horizon 2020 research and innovation programme under the Marie-Sklodowska-Curie grant agreement No. 794393.
We acknowledge the Virgo Consortium for making their simulation data available. The \eagle\ simulations were performed using the DiRAC-2 facility
at Durham, managed by the ICC, and the PRACE facility Curie based in France at TGCC, CEA, Bruy\`{e}res-le-Ch\^{a}tel.

\bibliographystyle{mnras}

\bibliography{Artale_GW}

\IfFileExists{\jobname.bbl}{}
{\typeout{}
\typeout{****************************************************}
\typeout{****************************************************}
\typeout{** Please run "bibtex \jobname" to optain}
\typeout{** the bibliography and then re-run LaTeX}
\typeout{** twice to fix the references!}
\typeout{****************************************************}
\typeout{****************************************************}
\typeout{}
}

\end{document}